\documentclass{aa}
\pdfoutput=1

\usepackage[intlimits]{amsmath}
\usepackage{txfonts}
\usepackage[american]{babel}
\usepackage{graphicx}
\usepackage{xspace}
\usepackage{natbib}

\newcommand{\Alfven}{Alfv\'{e}n\xspace}

\newcommand{\degree}{\ensuremath{^\circ}}
\newcommand{\unit}[1]{\ensuremath{\,\mathrm{#1}}}
\newcommand{\MURaM}{\texttt{MURaM}\xspace}

\newcommand{\lci}{\ensuremath{\lambda_\mathrm{ci}}}

\newcommand{\taus}{\ensuremath{\tau_\mathrm{s}}}

\begin{document}

\title{Vortices, shocks, and heating in the solar photosphere:\\ 
       effect of a magnetic field}
\author{R. Moll \and  R. H. Cameron\ \and M. Sch{\"u}ssler}
\institute{Max-Planck-Institut f\"{u}r Sonnensystemforschung, 
           Max-Planck-Stra{\ss}e 2, 37191 Katlenburg-Lindau, Germany}
\date{Accepted for publication in A\&A}
\abstract
{}
  {We study the differences between non-magnetic and magnetic regions in
  the flow and thermal structure of the upper solar photosphere. }
  {Radiative MHD simulations representing a quiet region and a plage
  region, respectively, which extend into the layers around the
  temperature minimum, are analyzed.}
  {The flow structure in the upper photospheric layers of the two
  simulations is considerably different: the non-magnetic
  simulation is dominated by a pattern of moving shock fronts while the
  magnetic simulation shows vertically extended vortices associated with
  magnetic flux concentrations. Both kinds of structures induce
  substantial local heating. The resulting average temperature profiles
  are characterized by a steep rise above the temperature minimum due to
  shock heating in the non-magnetic case and by a flat photospheric
  temperature gradient mainly caused by Ohmic dissipation in the
  magnetic run.}
  {Shocks in the quiet Sun and vortices in the strongly magnetized
    regions represent the dominant flow structures in the layers around
    the temperature minimum. They are closely connected with
    dissipation processes providing localized heating.}
\authorrunning{Moll et al.}
\titlerunning{Vortices, shocks, and heating in the solar photosphere}
\maketitle

\section{Introduction}

Semi-empirical models of strongly magnetized regions in the solar
atmosphere (such as network and plage areas) indicate increased
temperatures relative to the `quiet' Sun
\citep[e.g.][]{Fontenla:etal:2006}. This is consistent with the fact
that individual magnetic structures are hotter than their surroundings
in the upper layers of the photosphere and in the chromosphere
\citep[e.g.][]{Solanki:1993, Lagg:etal:2010}, but the nature of the
concomitant heating process is still under debate. One possibility is
the dissipation of mechanical energy in the form of flows or waves,
another is resistive dissipation of magnetic energy, for instance
in current sheets.

The dynamics in these atmospheric layers is complex and diverse. For the
weakly magnetized internetwork regions, numerical simulations and
observations \citep[e.g.][]{Carlsson:Stein:1992, Carlsson:Stein:1997,
Wedemeyer:etal:2004, Wedemeyer:2010} suggest that
shocks driven by overshooting convection play an important part above
the temperature-minimum region. In addition, swirling motions have
been detected in observations of the chromosphere
\citep{Wedemeyer:Rouppe:2009}, the transition region
\citep{Curdt:etal:2012}, and the corona \citep{Zhang:Liu:2011}.

Small-scale vortices occuring in simulations of non-magnetic or weakly
magnetized regions of the lower atmosphere and uppermost convection zone
were studied by \citet{Moll:etal:2011b} and
\citet{Kitiashvili:etal:2011b}. Simulations comprising sufficient
magnetic flux to represent network patches or plage regions exhibit
vertically orientated vortices associated with magnetic flux
concentrations in the upper photosphere and lower chromosphere
\citep{Voegler:2004a, Shelyag:etal:2011, Carlsson:etal:2010,
Kitiashvili:etal:2012, Steiner:Rezaei:2012}.

In this paper, we analyze simulations extending higher into the lower
chromosphere for a detailed study of the effect of a magnetic field on
the flow structure as well as on the heating processes in the atmosphere
up to a few hundred km above the average height of the temperature
minimum.  We compare a simulation without magnetic field as an extreme
case of `quiet Sun' with a simulation having a mean vertical field of
200~G, which is considered to represent the conditions in a plage
region.

\section{Methods}
\label{sec:methods}

\subsection{Simulations}
\label{subsec:simulations}

We analyzed simulations carried out with the \MURaM code \citep[for a
detailed description, see][]{Voegler:2003, Voegler:etal:2005}, which
treats the equations of compressible MHD together with the radiative
transfer equation solved along 12 rays per grid cell. The
equation of state incorporates the effect of partial ionization for 11
elements. The numerical scheme uses 4th-order centered spatial
differences and a 4th-order Runge-Kutta method for the time
stepping. Both the top and bottom boundaries permit free in- and outflow
of fluid, the side boundaries are periodic. The magnetic field at the
top and bottom boundaries is assumed to be vertical.

Here we consider two simulation runs, one non-magnetic and one
with an average unipolar vertical magnetic field of $200\unit{G}$.  Both
runs used a computational box with a horizontal area of
$6\times6\unit{Mm^2}$ and a height of $1.68\unit{Mm}$, ranging from
about 900~km below the average height of the optical surface to roughly
800~km above. The size of the grid cells was
$20.8\times20.8\times14.0\unit{km^3}$. The simulations were run for
several hours solar time to reach a statistically stationary state. For
the statistical analyses, we used 4 snapshots from each run.

\begin{figure*}[t]
\includegraphics[width=\linewidth]{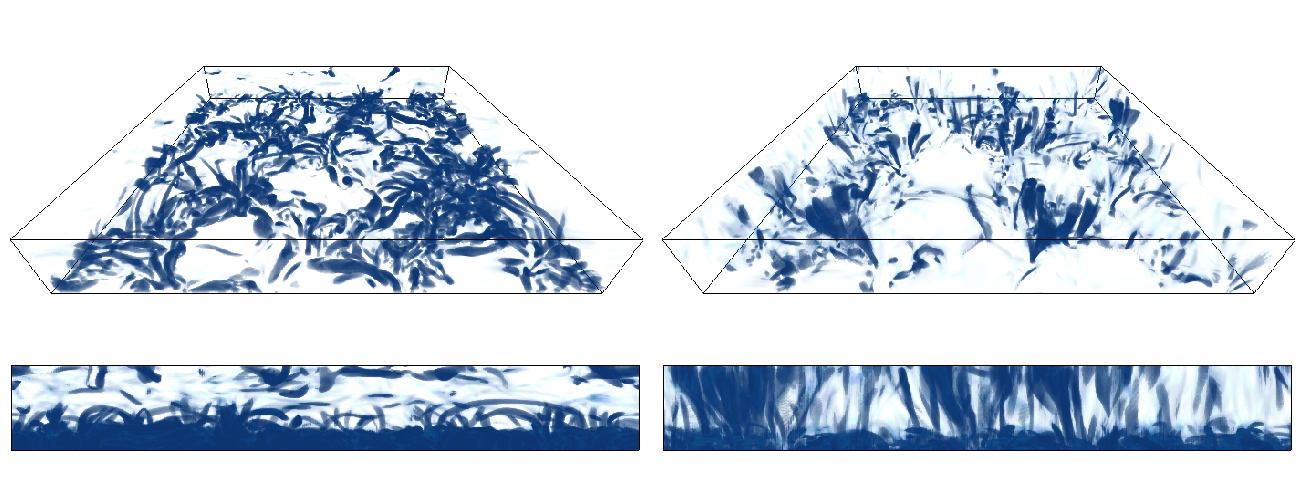}
\caption{Volume rendering of the swirling strength, $\lci$, in the
non-magnetic case (left) and the magnetic case (right)
from two viewing angles.  Shown is the upper half
($6\times6\times0.8\unit{Mm^3}$) of the computational domain, the bottom
plane corresponding roughly to the optical surface. The color scale
covers the range $0\le \lci \le 0.027\,$s$^{-1}$; higher values of
$\lci$ are saturated.}
\label{fig:omim3d}
\end{figure*}

\subsection{Vortex identification}
\label{sec:vortex}

The vorticity is not the optimal quantity to identify swirling flows.
For instance, a purely bidirectional shear flow without any rotational
component nevertheless has a non-vanishing vorticity. For the detection
of vortices, i.e., fluid elements rotating about a local, possibly
moving axis, we followed the procedure employed in
\citet{Moll:etal:2011b}: vortices (swirling flows) are defined as
regions where the velocity gradient tensor has a pair of complex
conjugate eigenvalues \citep{Zhou:etal:1999}. The strength of the
vortical motion is determined by the {\em swirling strength}, $\lci$,
which is the magnitude of the unsigned imaginary part of the complex
eigenvalues.  In the case of rigid rotation, we have $\lci=2\pi/\taus$,
where $\taus$ is the rotation (or swirling) period.  The direction of a
vortex is defined to be along the eigenvector corresponding to the real
eigenvalue.  The inclination angle of this vector with respect to the
vertical direction is denoted with $\iota$.

\section{Results}
\label{sec:results}

\subsection{Vortex properties}
\label{subsec:vortices}

\begin{figure*}[ht!]
\centering
\includegraphics[width=0.6\linewidth]{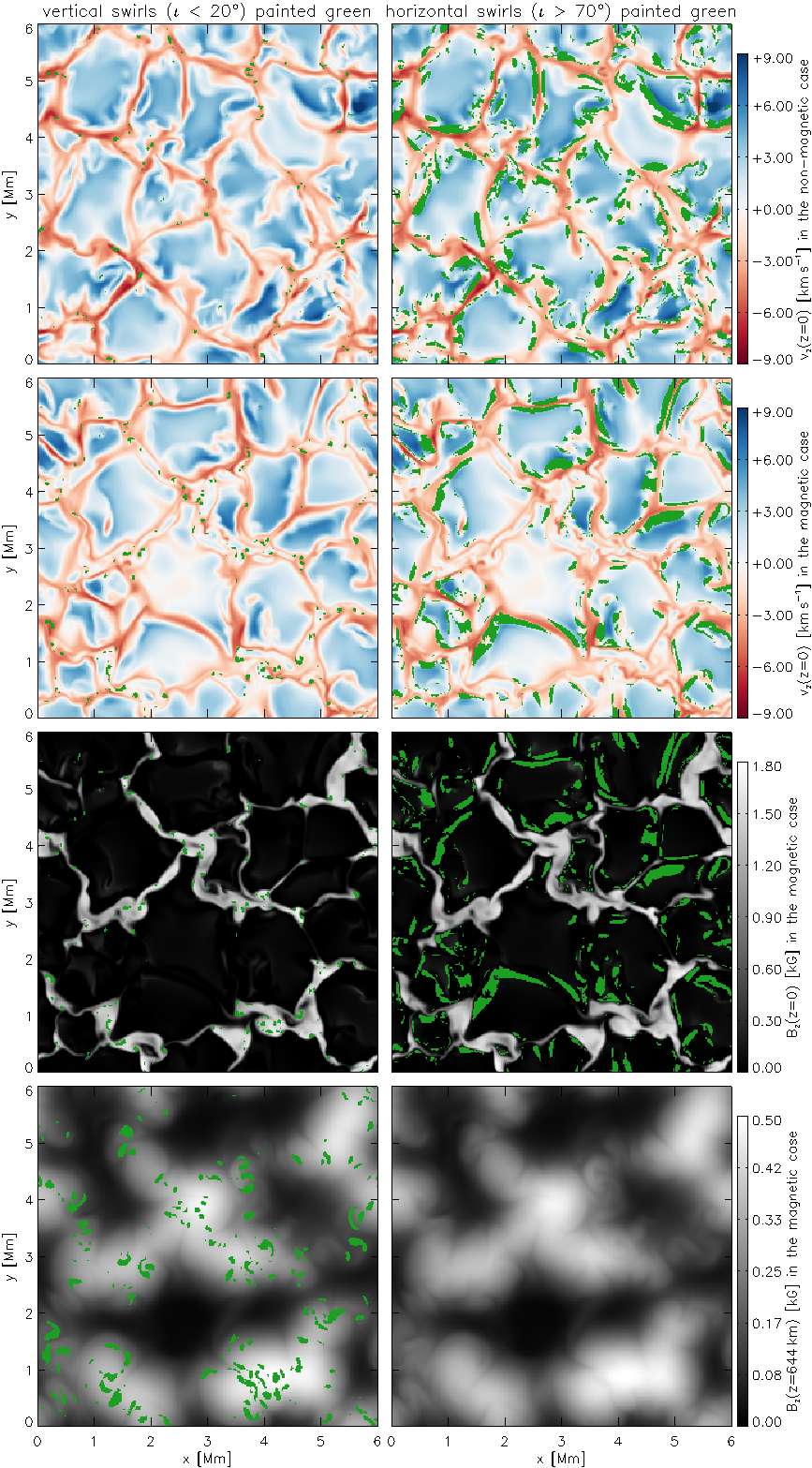}
\caption{Distribution of vertical vortices (inclination $\iota <
  20^\circ$, left column) and horizontal vortices ($\iota > 70^\circ$,
  right column) on horizontal planes. Pixels corresponding to swirling
  periods $\tau_s < 240\,\mathrm{s}$ are indicated in green color on
  cuts of the vertical velocity at $z=0$ (average height of the optical
  surface) for the non-magnetic case (top row) and for the magnetic case
  (second row). Similarly, the strong swirls are indicated on maps of
  the vertical magnetic field component at $z=0$ (third row) and at
  $z=644\,$km (bottom row) for the simulation with magnetic field.}
\label{fig:vortex_locs}
\end{figure*}

Figure~\ref{fig:omim3d} shows regions of high swirling strength in
snapshots from the non-magnetic and magnetic simulations,
respectively. The most striking difference between the two cases is the
dominance of tall, vertically orientated vortices in the magnetic case,
which extend over the entire height of the simulated photosphere. In the
non-magnetic case, the vortex features do not protrude far above the
optical surface, bending over to form mainly horizontally orientated
loops. Vortex features near the top of the computational domain are
associated with shock fronts (see Sec.~\ref{subsec:dynamics}). The
difference between the two simulations is further illustrated in
Fig.~\ref{fig:vortex_locs}, which shows the location of vertical ($\iota
< 20^\circ$, left column) and horizontal ($\iota > 70^\circ$, 
right column) vortices on horizontal planes. In the upper two rows, the
vortex locations are compared to the distribution of vertical velocity
at the average height of the optical surface $(z=0)$. Near the optical
surface, both simulations show horizontal vortices at the edges of
granules and vertical vortices in the intergranular downflows lanes. The
comparison with the magnetic-field distribution in the third row
indicates that the horizontal vortices near $z=0$ are a hydrodynamical
phenomenon of the overturning motions at the borders of granules, which
is essentially unrelated to the intergranular magnetic flux
concentrations \citep[cf.][]{Steiner:etal:2010, Moll:etal:2011b}.  In
contrast to this, the vertical vortices are almost exclusively found in
magnetic flux concentrations. This one-to-one relationship extends
throughout the atmosphere, as shown in the bottom row of
Fig.~\ref{fig:vortex_locs}, which corresponds to a height of $z=644\,$km
above the average optical surface. At this height, we find a multitude
of vertical vortices in the (expanded) flux concentrations, while strong
horizontal vortices are absent.

The visual impression provided by Figs.~\ref{fig:omim3d} and
\ref{fig:vortex_locs} is confirmed quantitatively in Fig.~\ref{fig:lps},
which shows height profiles of horizontally averaged vorticity, swirling
strength, and the shear part of vorticity, $\omega_{\rm sh}= \omega -
2\lci$, where $\omega$ is the modulus of the vorticity vector.%
\footnote{For a purely rotational flow, we have
$\omega=2\lci=\omega_{\rm R}$ at the center of the vortex. Therefore,
one can formally split the vorticity into a rotational part,
$\omega_{\rm R}$, and the remainder, $\omega_{\rm sh}=\omega -
\omega_{\rm R}$, which gives the contribution from locally non-rotating
shear flow.}
In the magnetic simulation, all three quantities are suppressed in the
deeper layers up to $z\simeq200\,$km, owing to the effect of the Lorentz
force on the convective motions. In the upper layers, the situation is
reversed and the swirling and shearing motions are enhanced in
comparison to the non-magnetic case. We surmise that this is caused by
the dynamical coupling of the vortex motion in the convective
near-surface layers to the upper layers, which is mediated by the mainly
vertical magnetic field.

The drastic change in the properties of the vortex flows above about
150~km height is analyzed in more detail in Fig.~\ref{fig:depthcov},
which shows height profiles of the fractions of horizontal area covered
by swirls of different strength and orientation. In the non-magnetic
case (upper panel), horizontal vortices dominate at all heights, but the
area fraction of strong vortices (swirling period $\taus<2\,$min) drops
drastically above the optical surface. For the vertical vortices, the
decrease starts even below the optical surface.  The increase of
horizontal vortices above about 500~km is probably related to the
pattern of shock waves developing in these layers (see
Sect.~\ref{subsec:dynamics}). In the magnetic case (lower panel of
Fig.~\ref{fig:depthcov}), the horizontal vortices behave very similar to
those in the non-magnetic simulation, their area coverage dropping even
faster with height. However, the area fraction of vertical vortices
shows almost no drop above $z=0$, followed by a strong increase above
200~km, from which height on they dominate. Virtually all swirls at
large heights have vertical orientation
($\iota\mathord{\lesssim}20\degree$) in the magnetic case.  Even very
strong vertical vortices with periods below one minute reach appreciable
area coverage above $z\simeq400\,$km.

\newcommand{\myscale}{.45}
\begin{figure}[ht!]
\centering
\includegraphics[scale=\myscale]{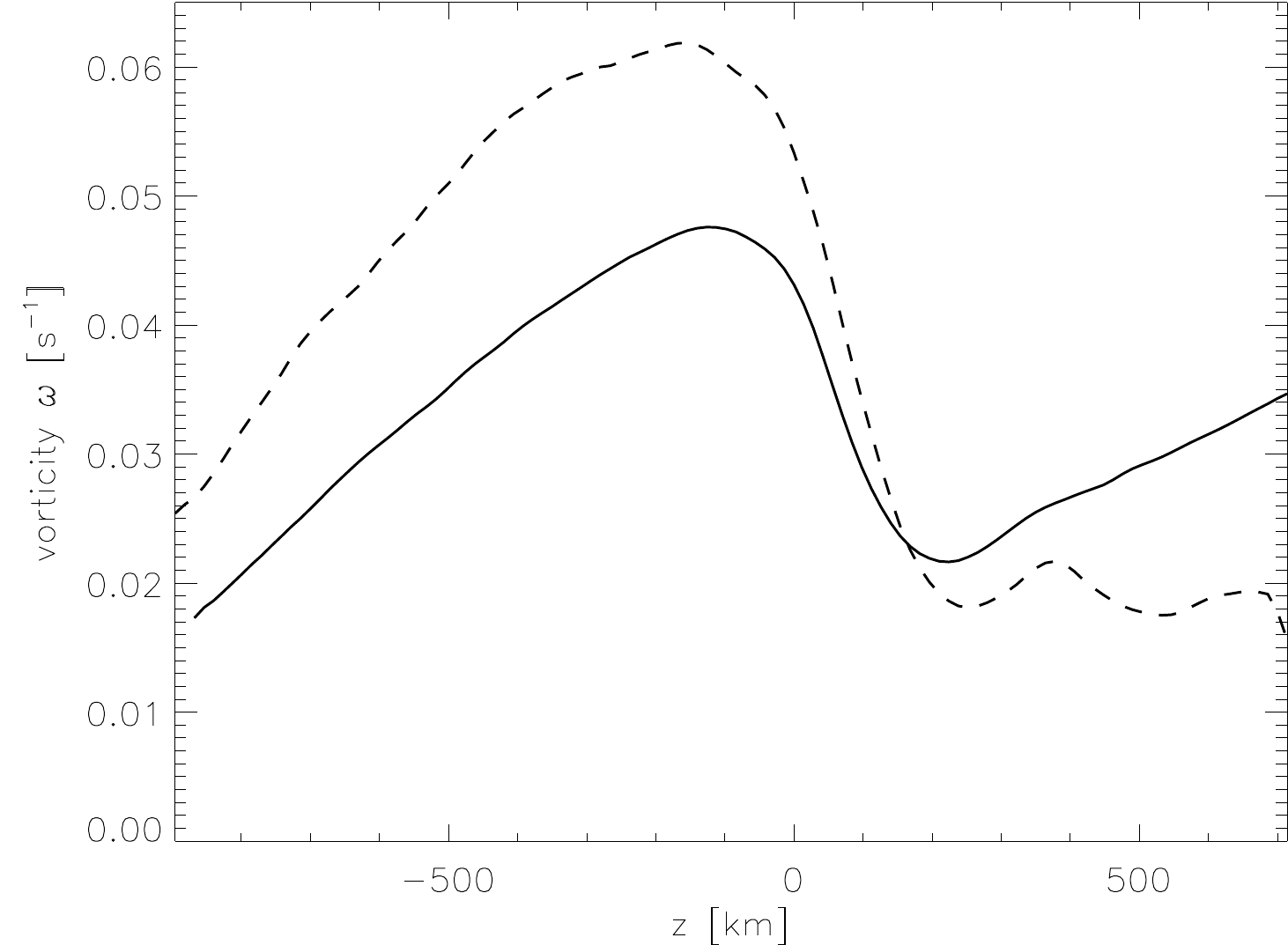} 
\vglue 3mm
\includegraphics[scale=\myscale]{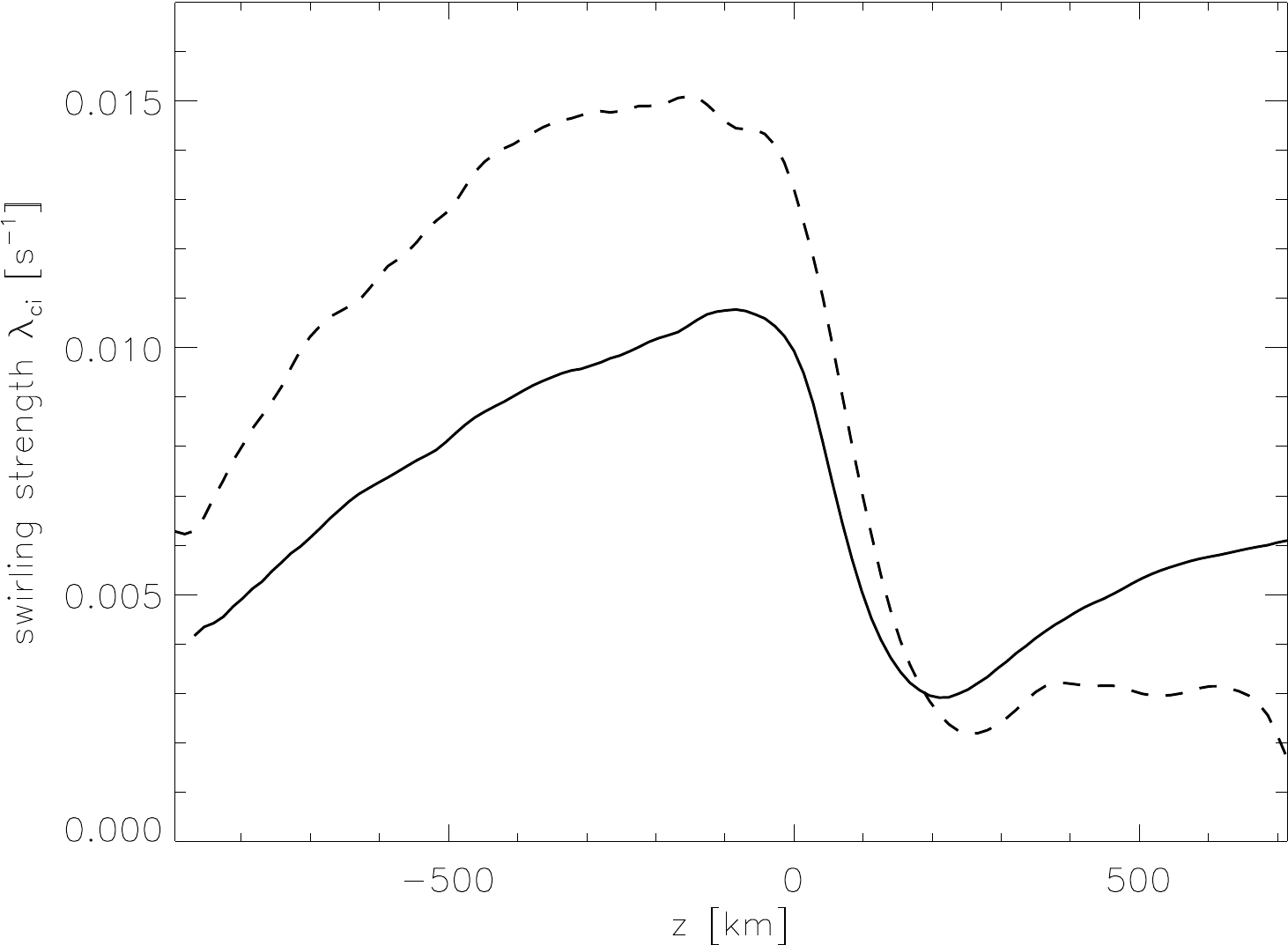}
\vglue 3mm
\includegraphics[scale=\myscale]{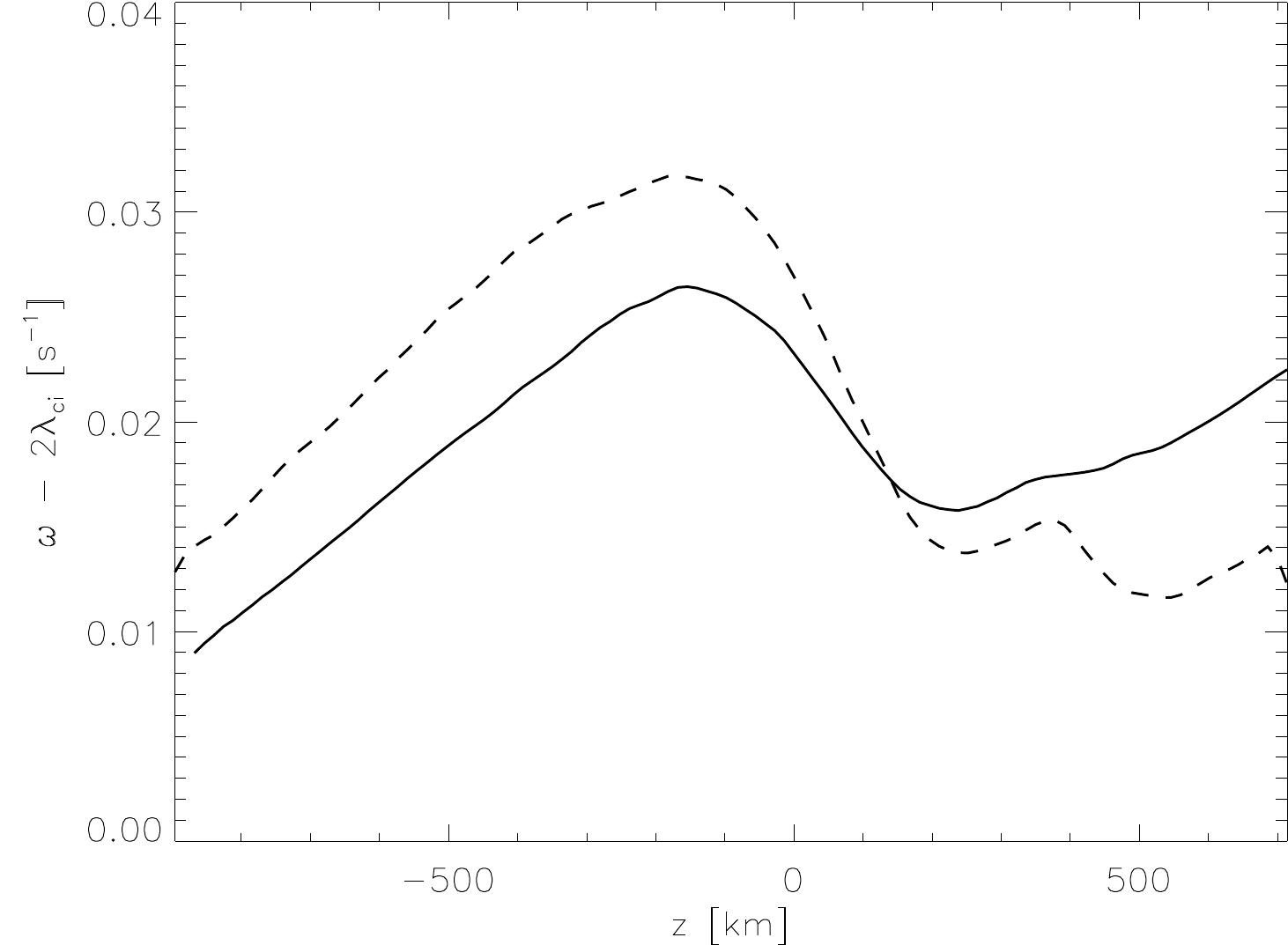} 
\caption{Horizontal averages of vorticity (top), swirling strength
 (middle), and shear part of the vorticity (bottom), as functions of
 height.  Solid lines correspond to the magnetic case, dashed lines to
 the non-magnetic case.}
\label{fig:lps}
\end{figure}

\renewcommand{\myscale}{.45}
\begin{figure}[ht!]
\centering
\includegraphics[scale=\myscale]{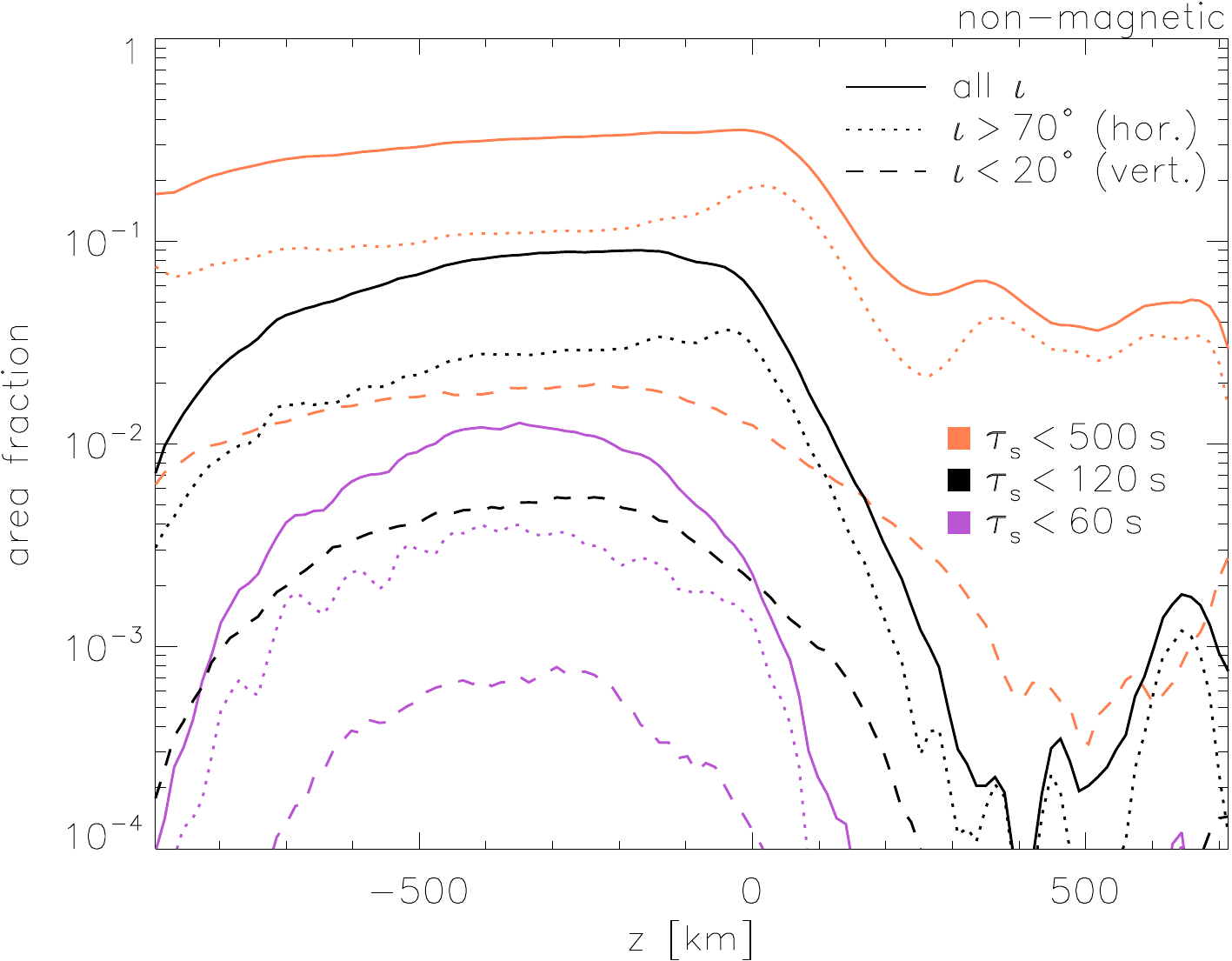} 
\\[\medskipamount]
\includegraphics[scale=\myscale]{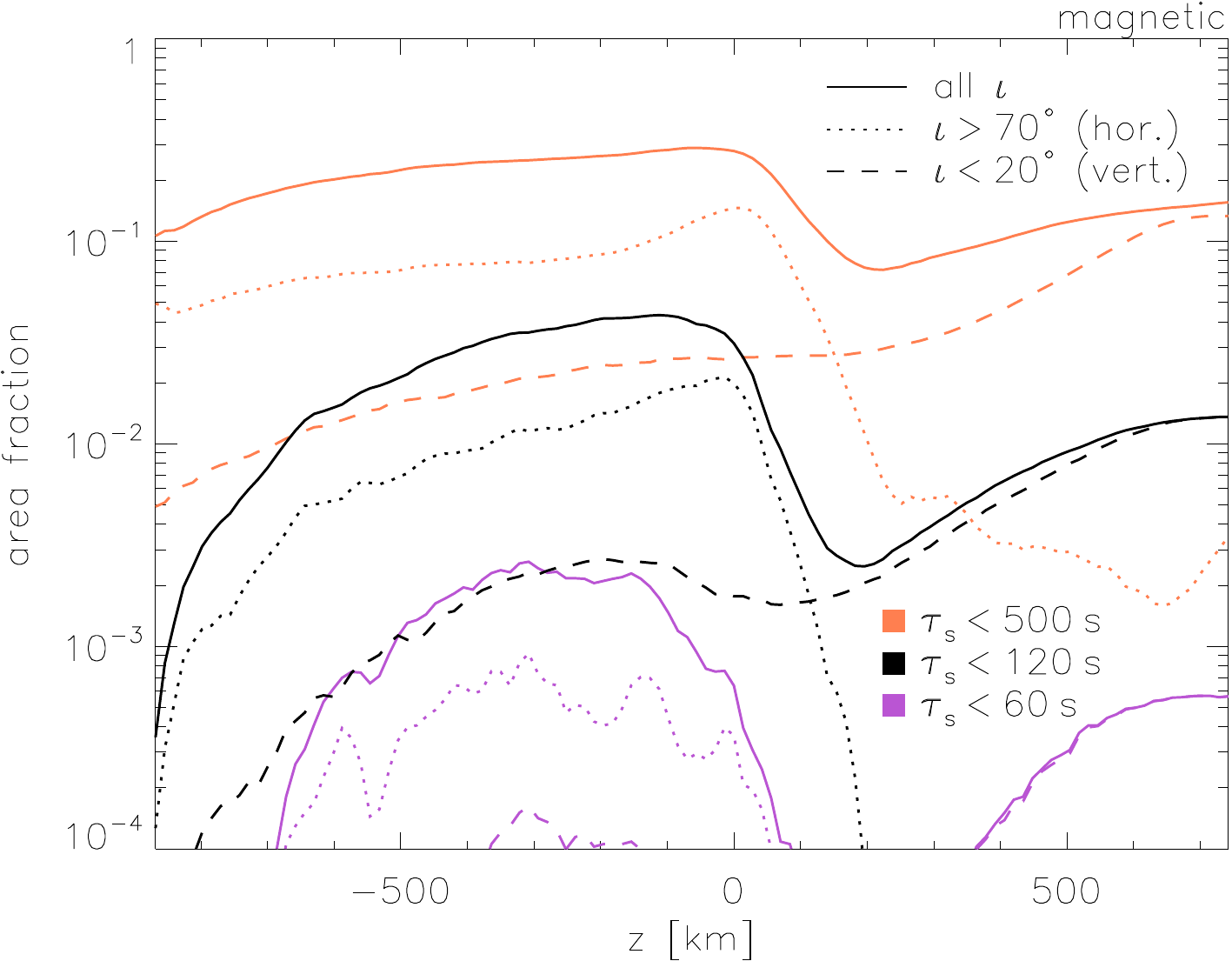}
\caption{Height profiles of the horizontal area fraction of grid cells
  with swirling period, $\taus=2\pi/\lci$, below given thresholds in the
  non-magnetic case (upper panel) and in the magnetic case (lower
  panel). Full lines correspond to vortices of all inclinations, while
  dotted and dashed lines separately represent horizontal and vertical
  swirls, respectively.}
\label{fig:depthcov}
\end{figure}

Fig.~\ref{fig:1Dhist} shows histograms of vortex properties at three
height ranges, one in the deep layers of the computational domain
(brown curves), one straddling the optical surface (green
curves), and one high in the photosphere (blue curves). While the
vortices in the non-magnetic case (dashed curves) and the magnetic
case (solid curves) do not strongly differ in the deep and the
near-surface layers, they show strong deviations in the higher layer: in
the magnetic run, the typical swirling periods are much smaller (i.e.,
the vortices are more vigorous) and the gas pressure is strongly reduced
in comparison to the average gas pressure at the same height. The low
pressure results from the location of the vortices in magnetic flux
concentrations (cf.  Fig.~\ref{fig:vortex_locs}, bottom left panel), 
together with the effect of the centrifugal force.  The temperatures of
the weak swirls in the upper photosphere of the non-magnetic simulation
are significantly higher than the average temperature, presumably due to
their association with shock fronts. The vortices in the upper layers of
the magnetic simulation are also hotter by 10--20\% compared to their
average surroundings. The vertical velocities in vortices do not differ
strongly between the two simulations: in the deep and near-surface
layers, vortices are predominantly located in downflows in both cases,
while in the upper layer there is a slight preference for upflows in the
non-magnetic case and an almost symmetric distribution around zero in
the magnetic run.

\renewcommand{\myscale}{.45}
\begin{figure*}[t!]
\centering
{\includegraphics[scale=\myscale]{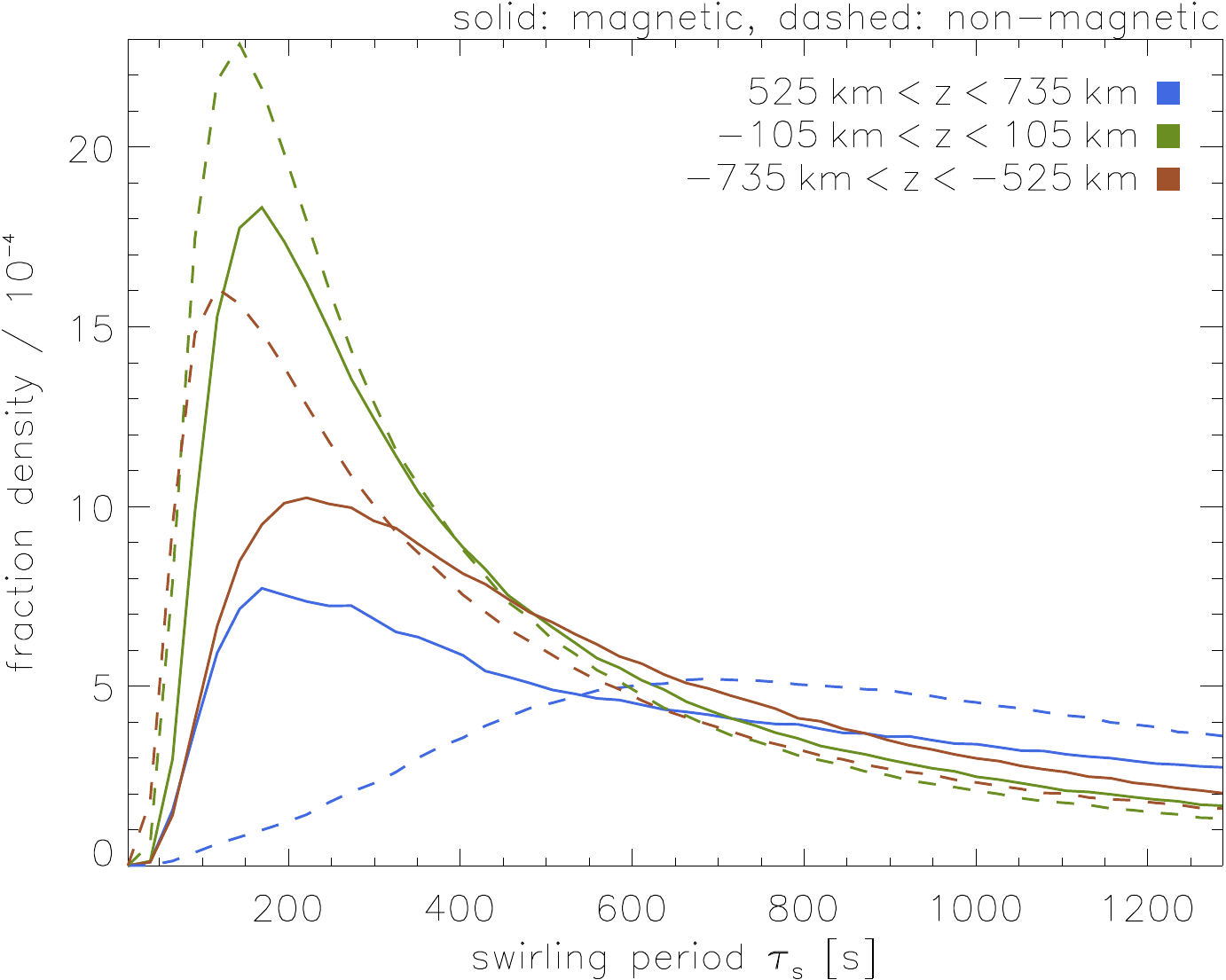}}\hglue 4mm 
{\includegraphics[scale=\myscale]{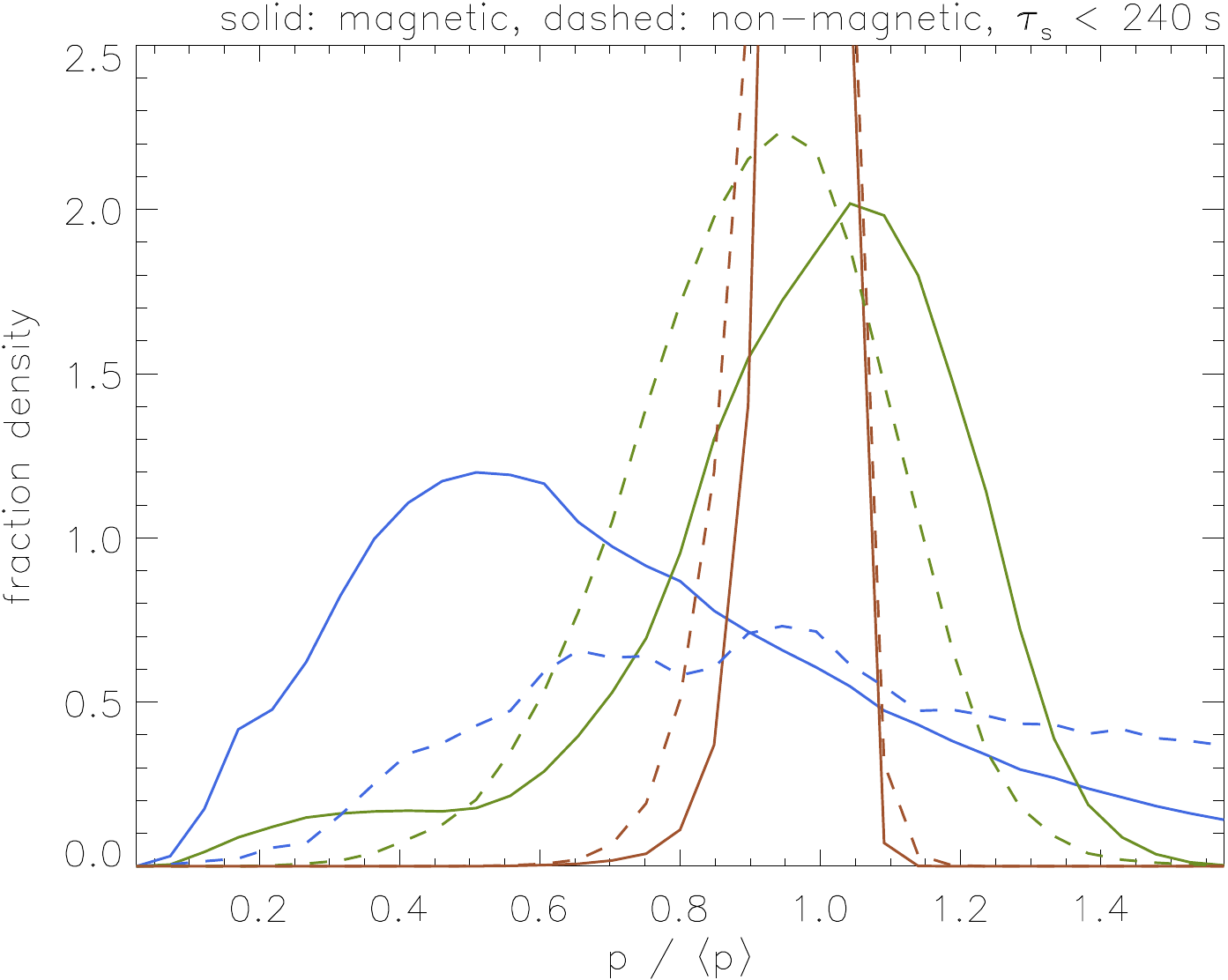}}
          \\[\medskipamount]
{\includegraphics[scale=\myscale]{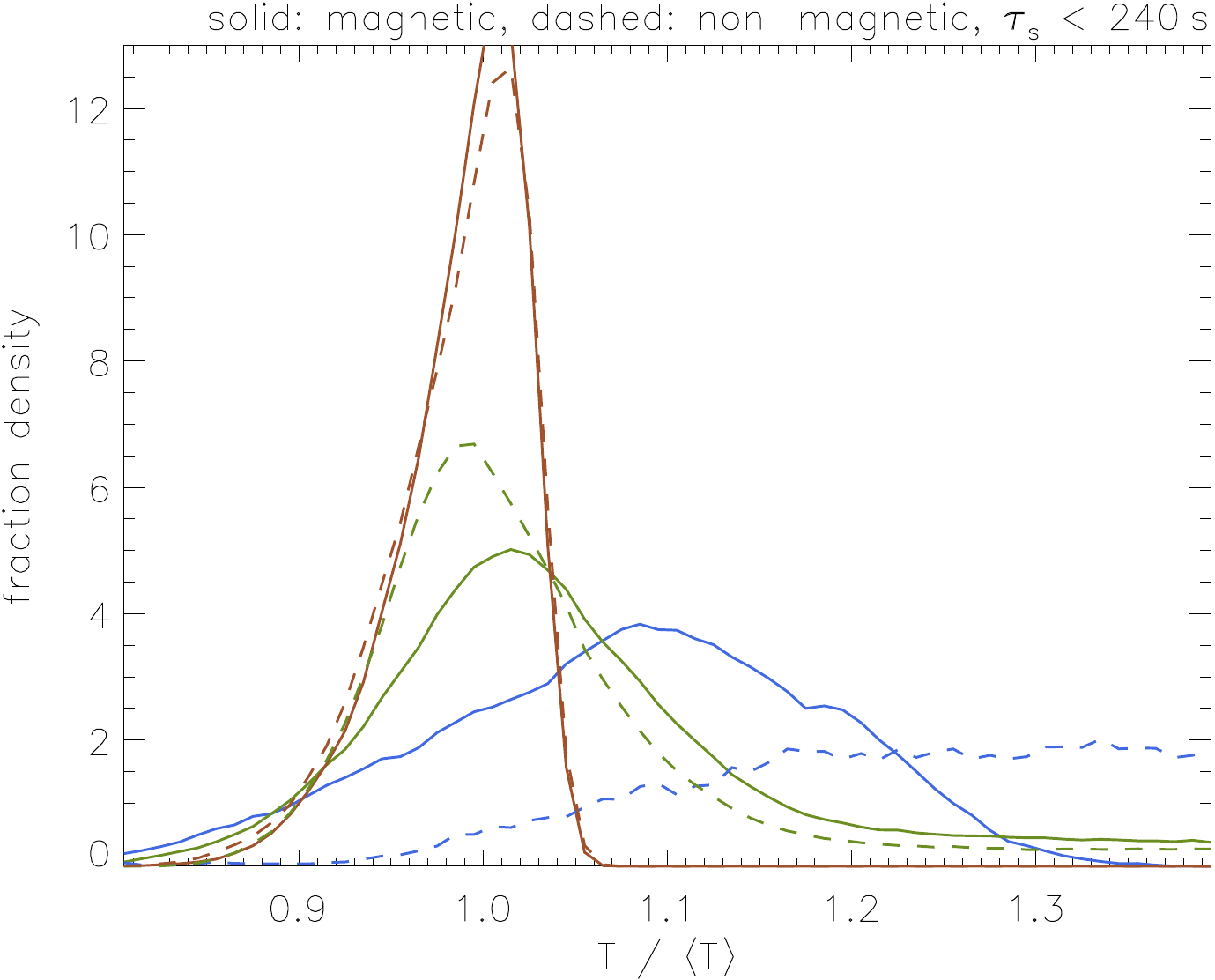}}\hglue 4mm
{\includegraphics[scale=\myscale]{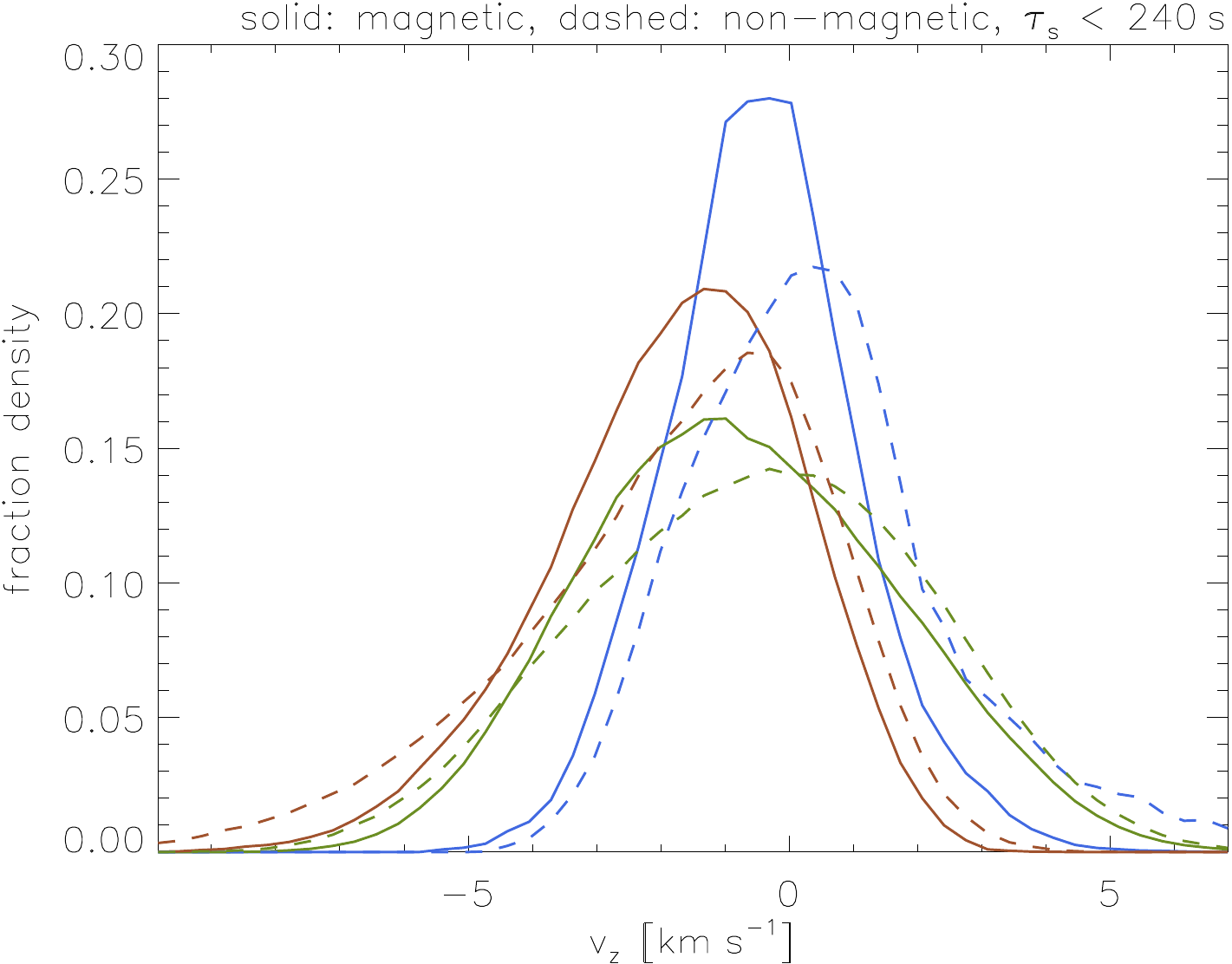}}
\caption{Histograms of various quantities in swirling regions for three
 layers of $210\,$km width each, centered at heights $z=-630\unit{km}$,
 $z=0\unit{km}$, and $z=630\unit{km}$. Solid lines represent the
 magnetic case, dashed lines the non-magnetic case. The quantities
 considered are swirling period (upper left panel), gas pressure
 normalized to its horizontal average (upper right), temperature
 normalized to its horizontal average (lower left), and vertical
 velocity component (lower right).  Each histogram is normalized by its
 integral.}
\label{fig:1Dhist}
\end{figure*}

\renewcommand{\myscale}{.5}
\begin{figure*}[ht!]
\centering
{\includegraphics[scale=\myscale]{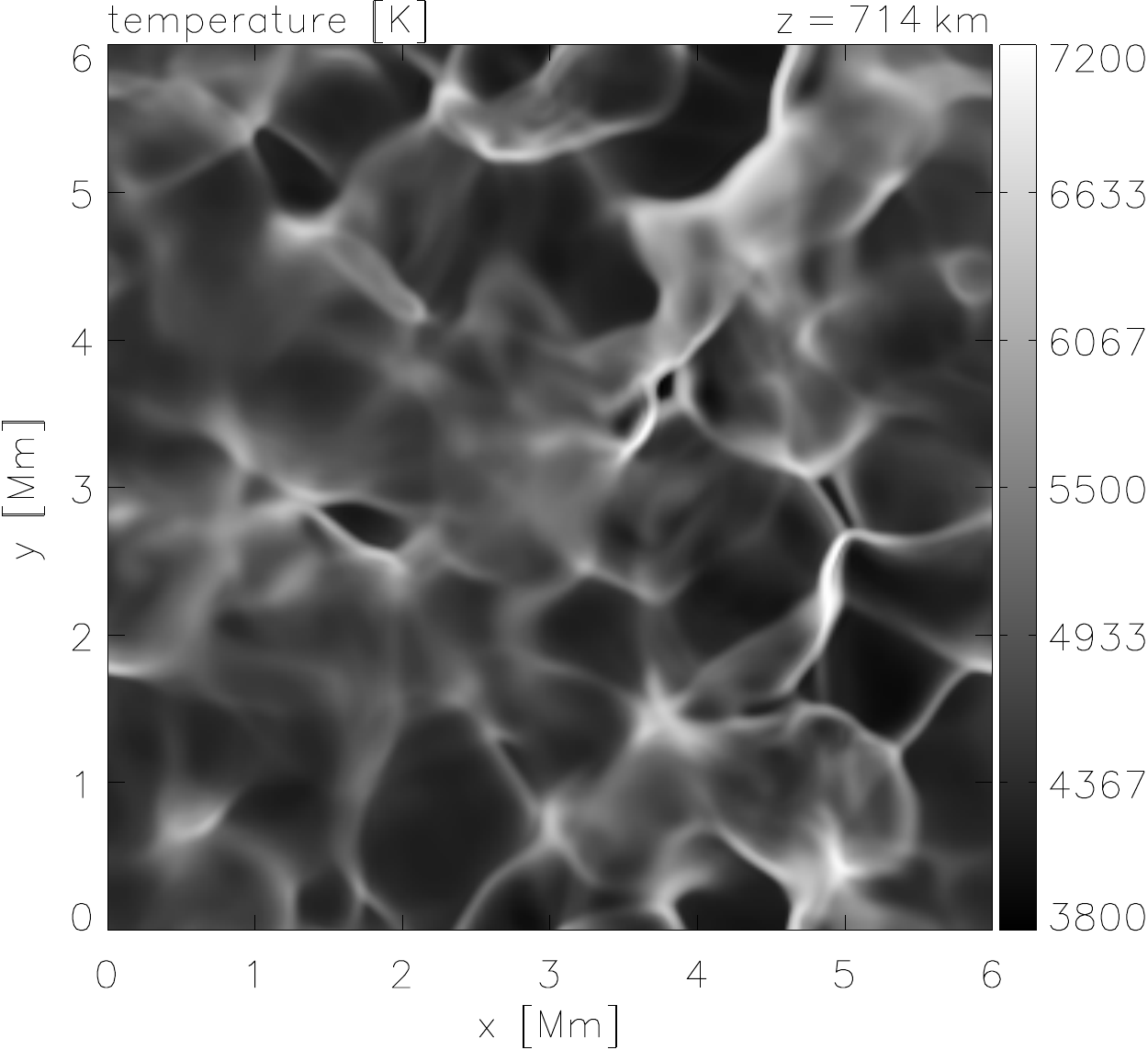}}\hglue 4mm
{\includegraphics[scale=\myscale]{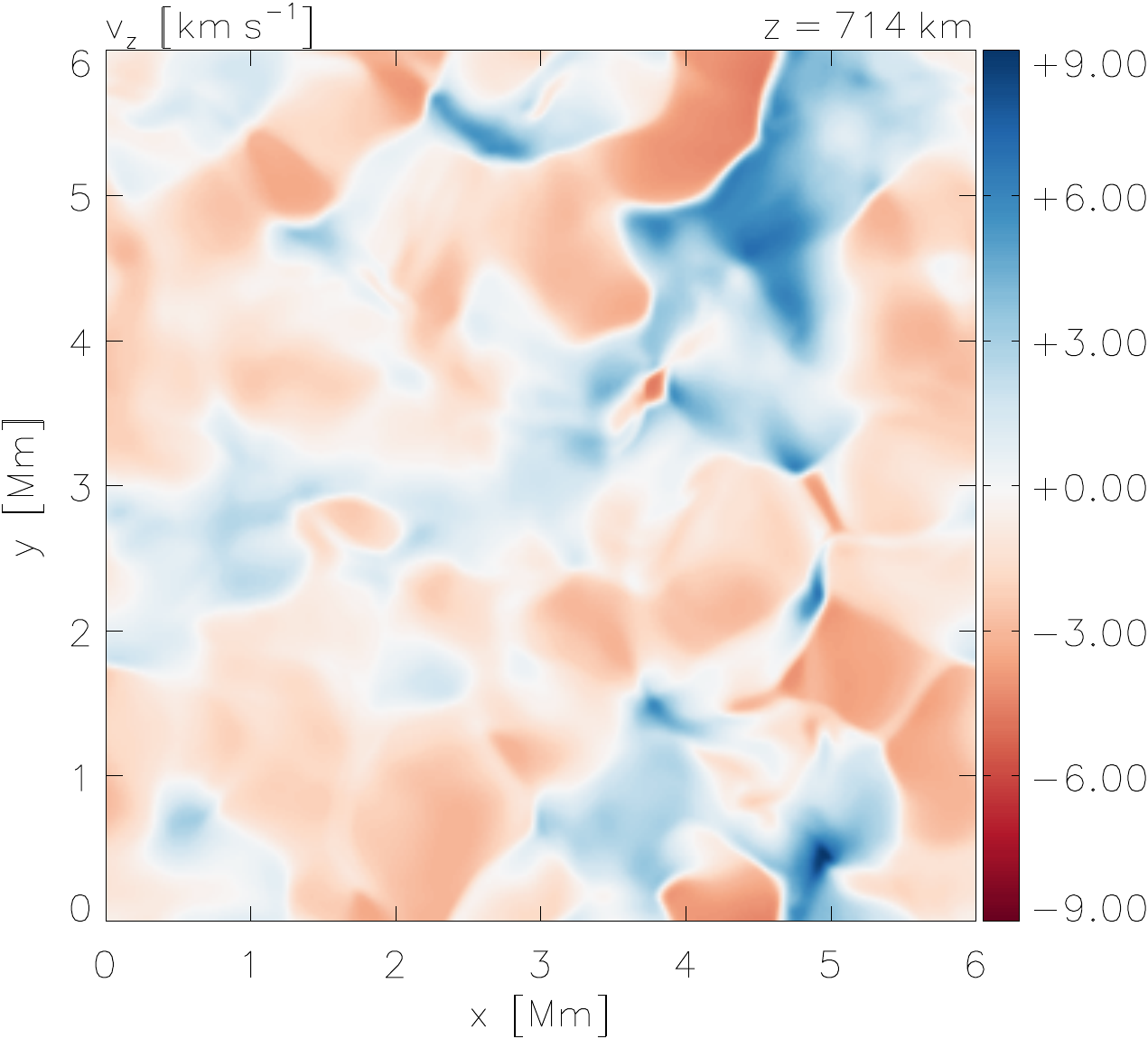}} 
 \\[\medskipamount]
{\includegraphics[scale=\myscale]{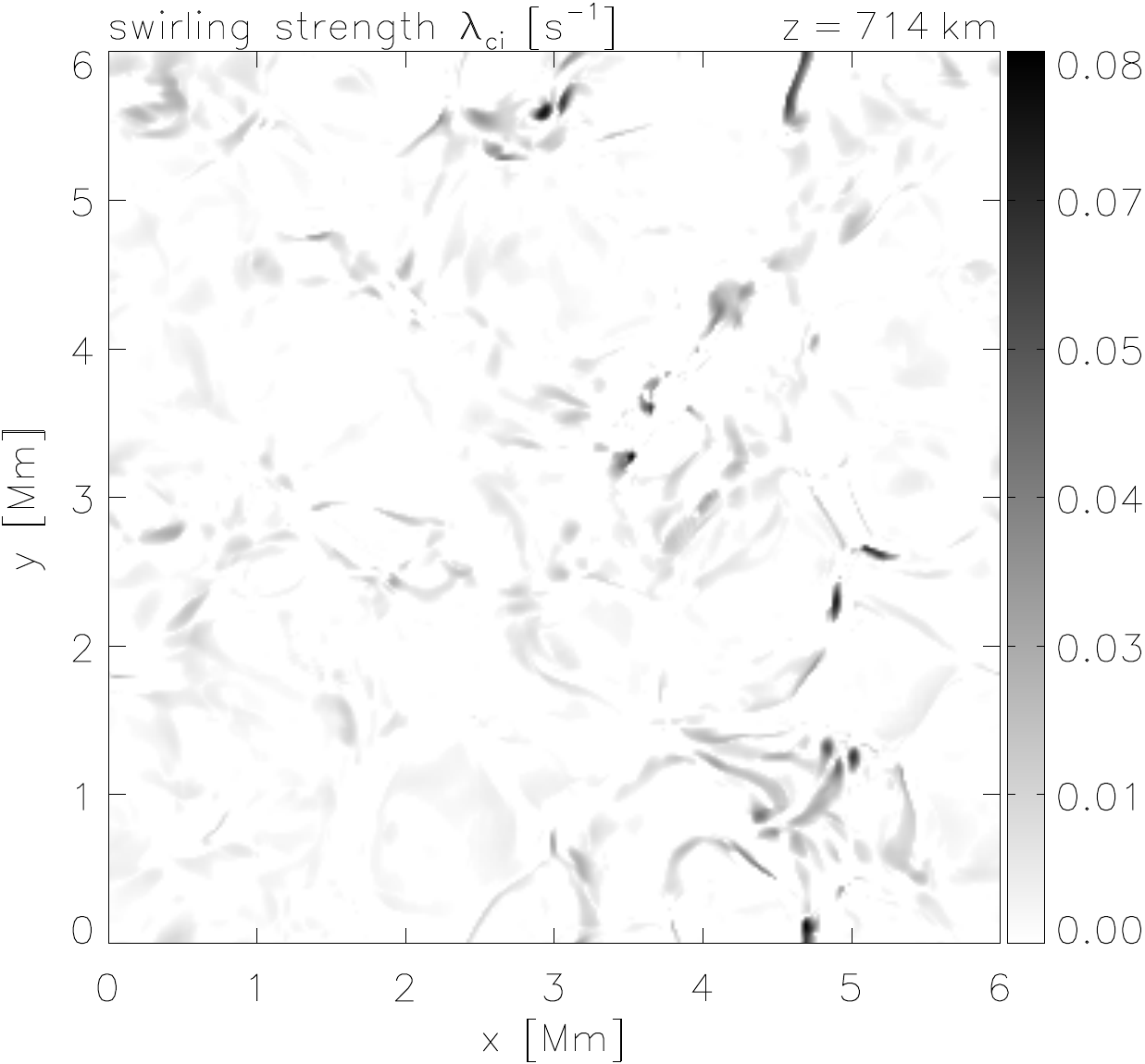}}\hglue 4mm
{\includegraphics[scale=\myscale]{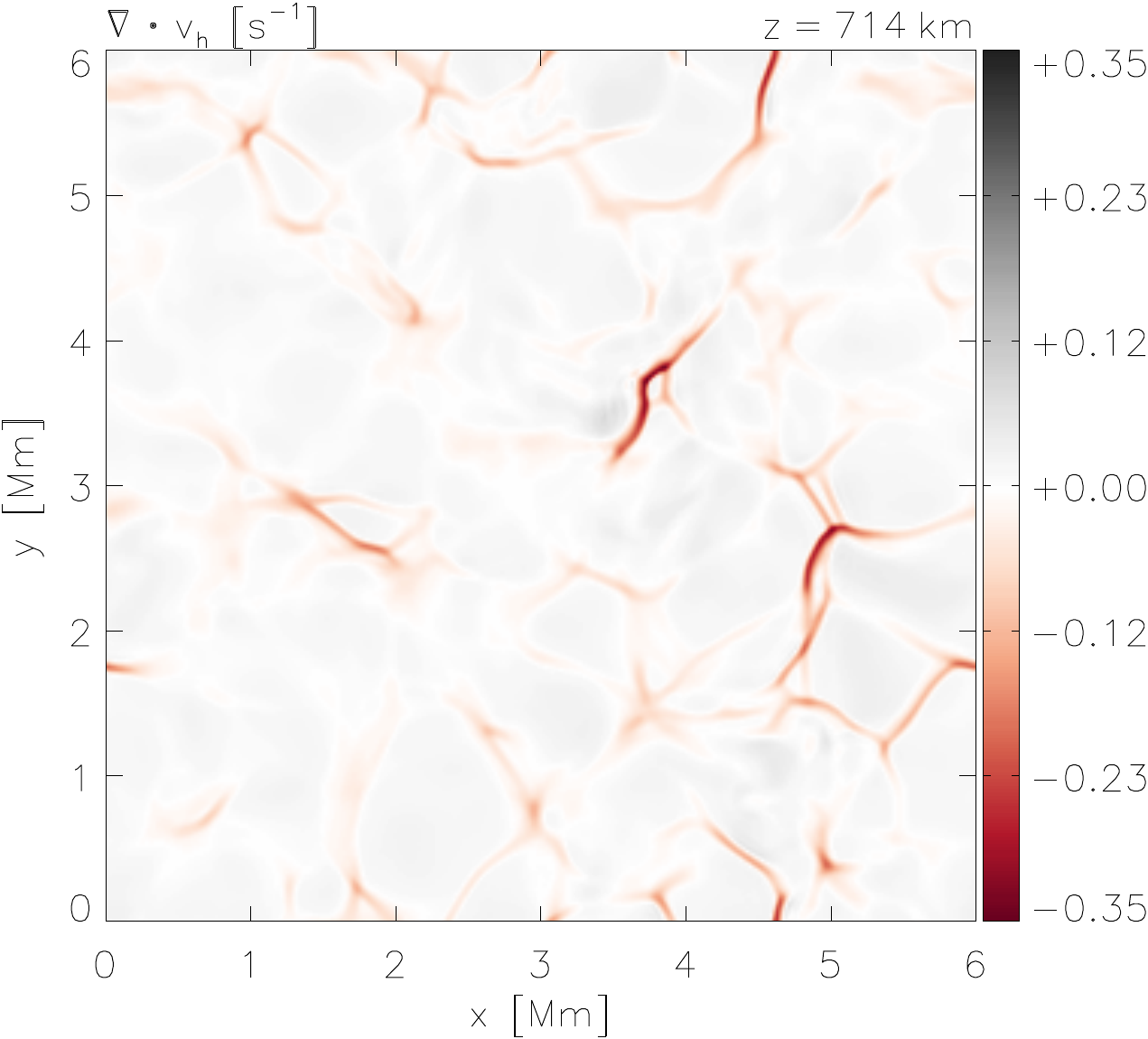}} 
 \\[\medskipamount]
{\includegraphics[scale=\myscale]{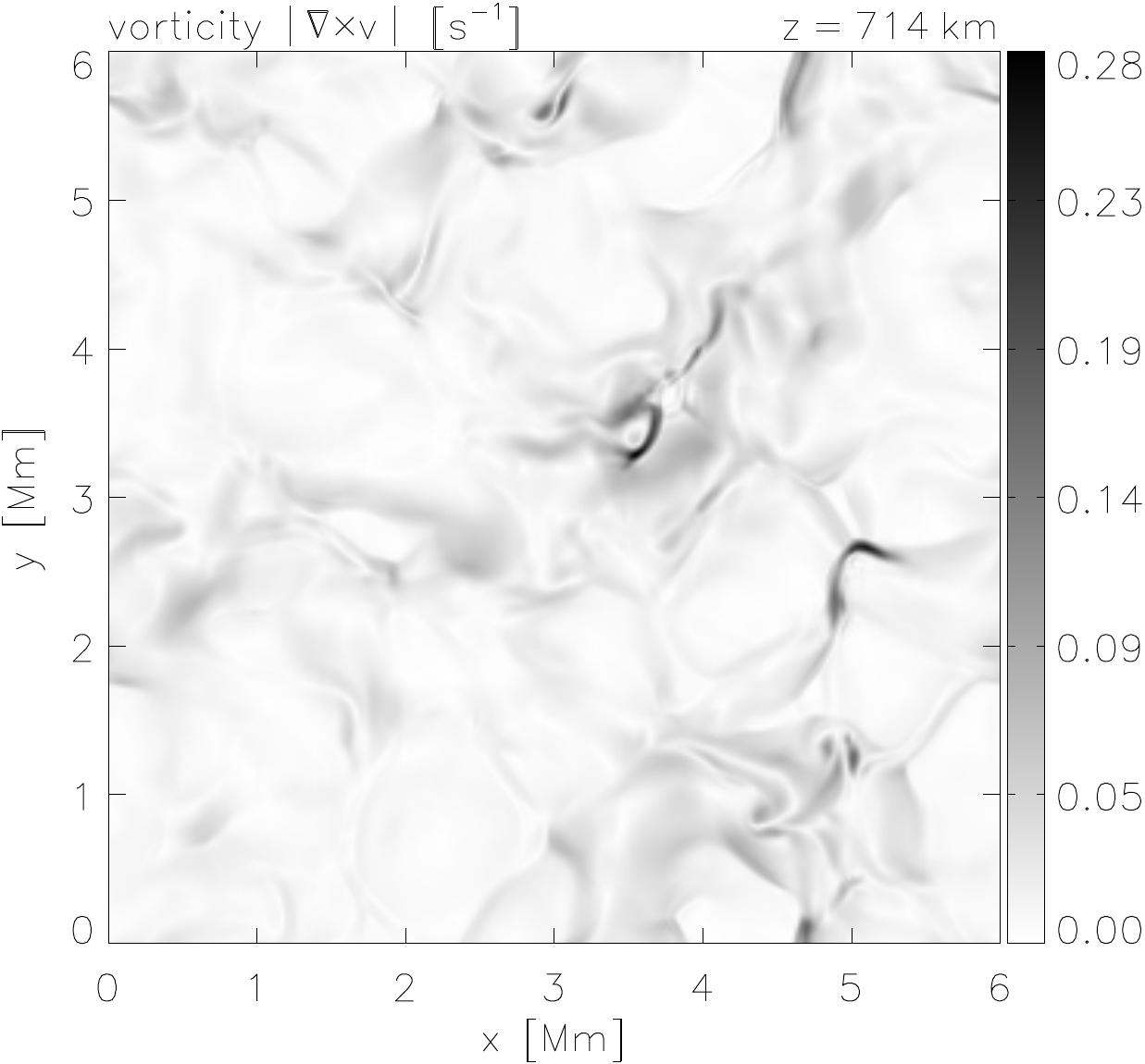}}\hglue 4mm
{\includegraphics[scale=\myscale]{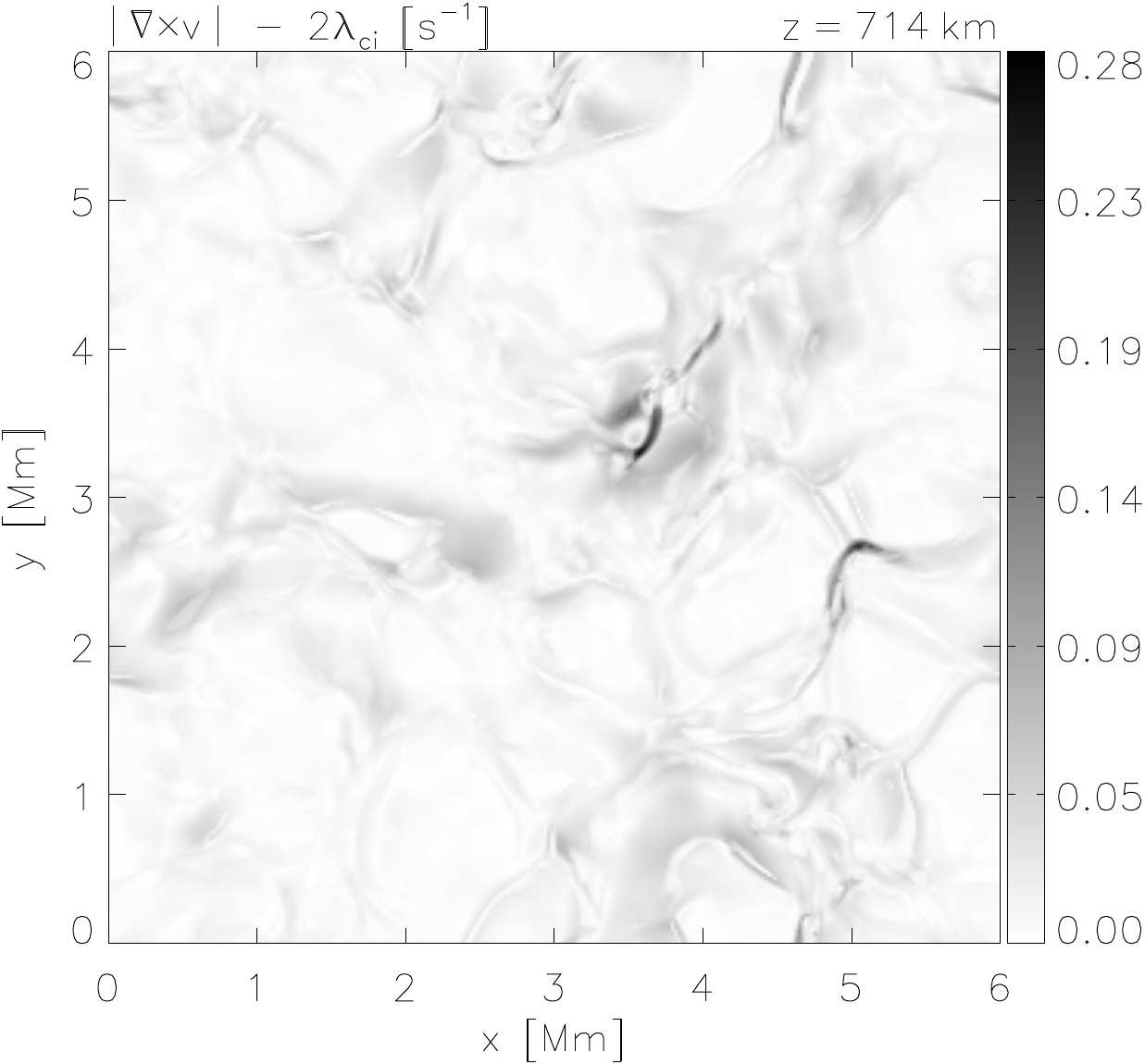}}
\caption{Horizontal maps of various quantities at $z=714\unit{km}$ for
 the non-magnetic case. {\em From left to right, top to bottom}:
 temperature, vertical velocity, swirling strength, divergence of the
 horizontal velocity, vorticity, and shear part of vorticity.}
\label{fig:mapsH}
\end{figure*}

\renewcommand{\myscale}{.5}
\begin{figure*}[pt!]
\centering
{\includegraphics[scale=\myscale]{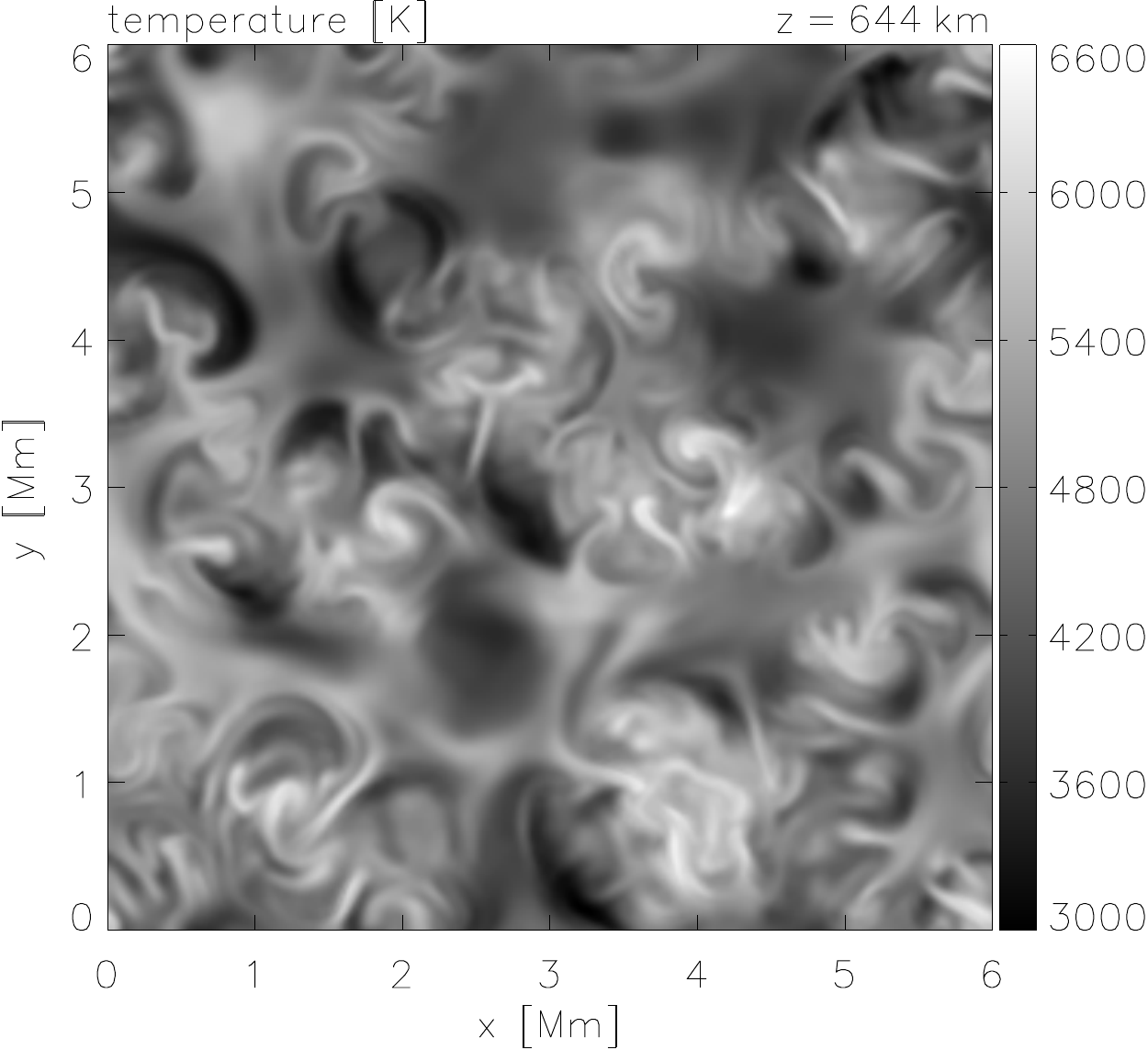}}\hglue 4mm
{\includegraphics[scale=\myscale]{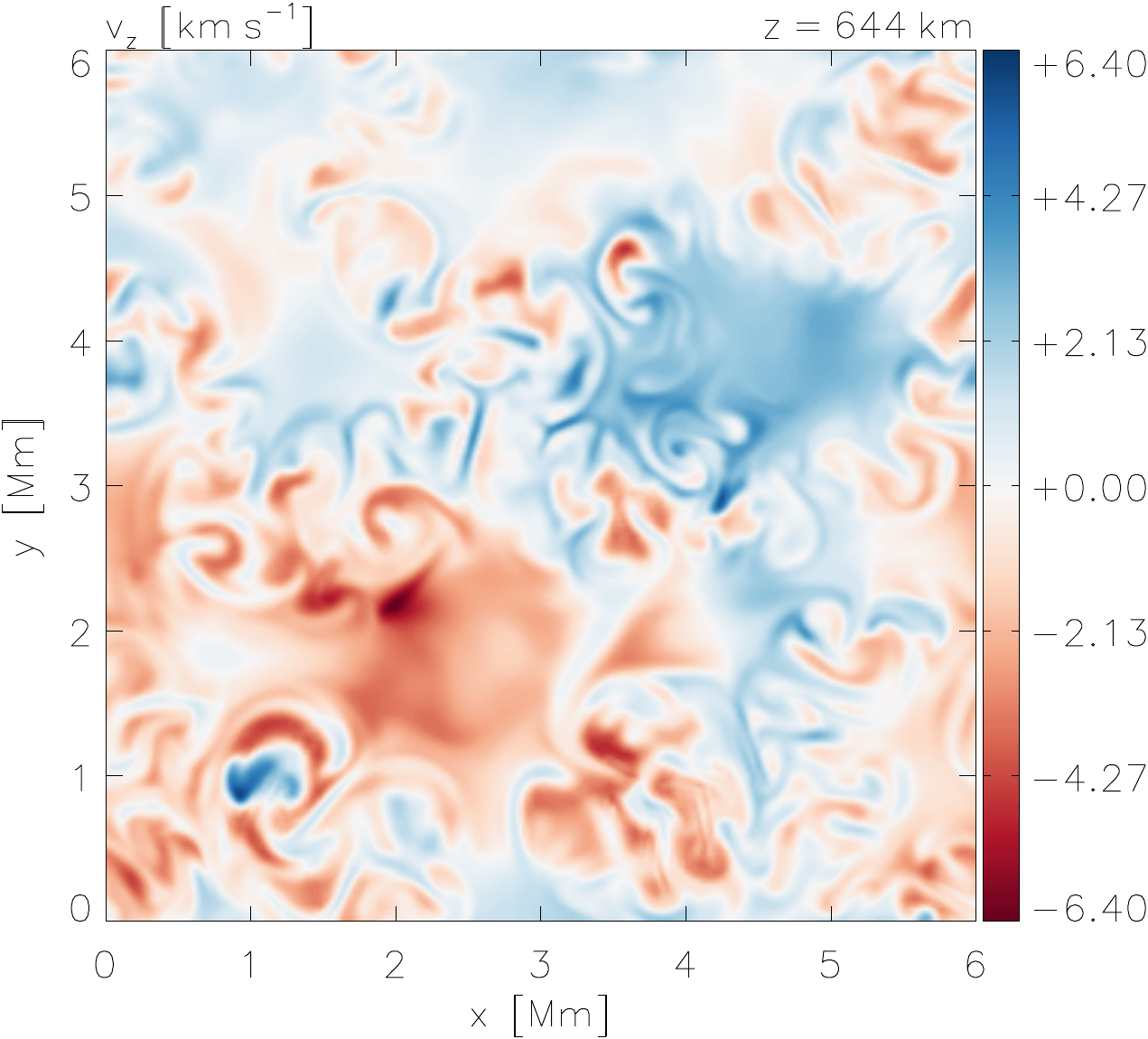}} 
          \\[\medskipamount]
{\includegraphics[scale=\myscale]{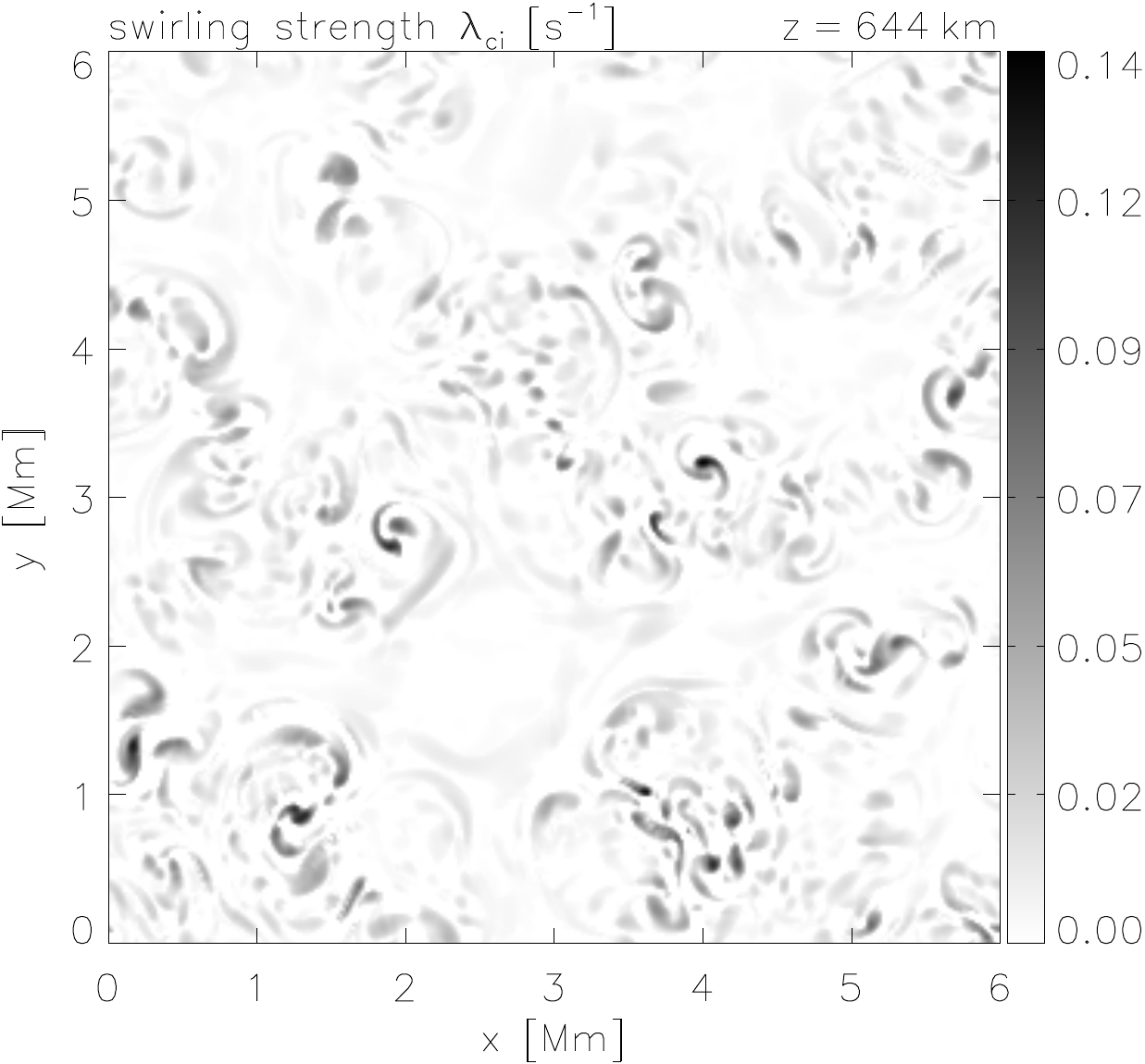}}\hglue 4mm
{\includegraphics[scale=\myscale]{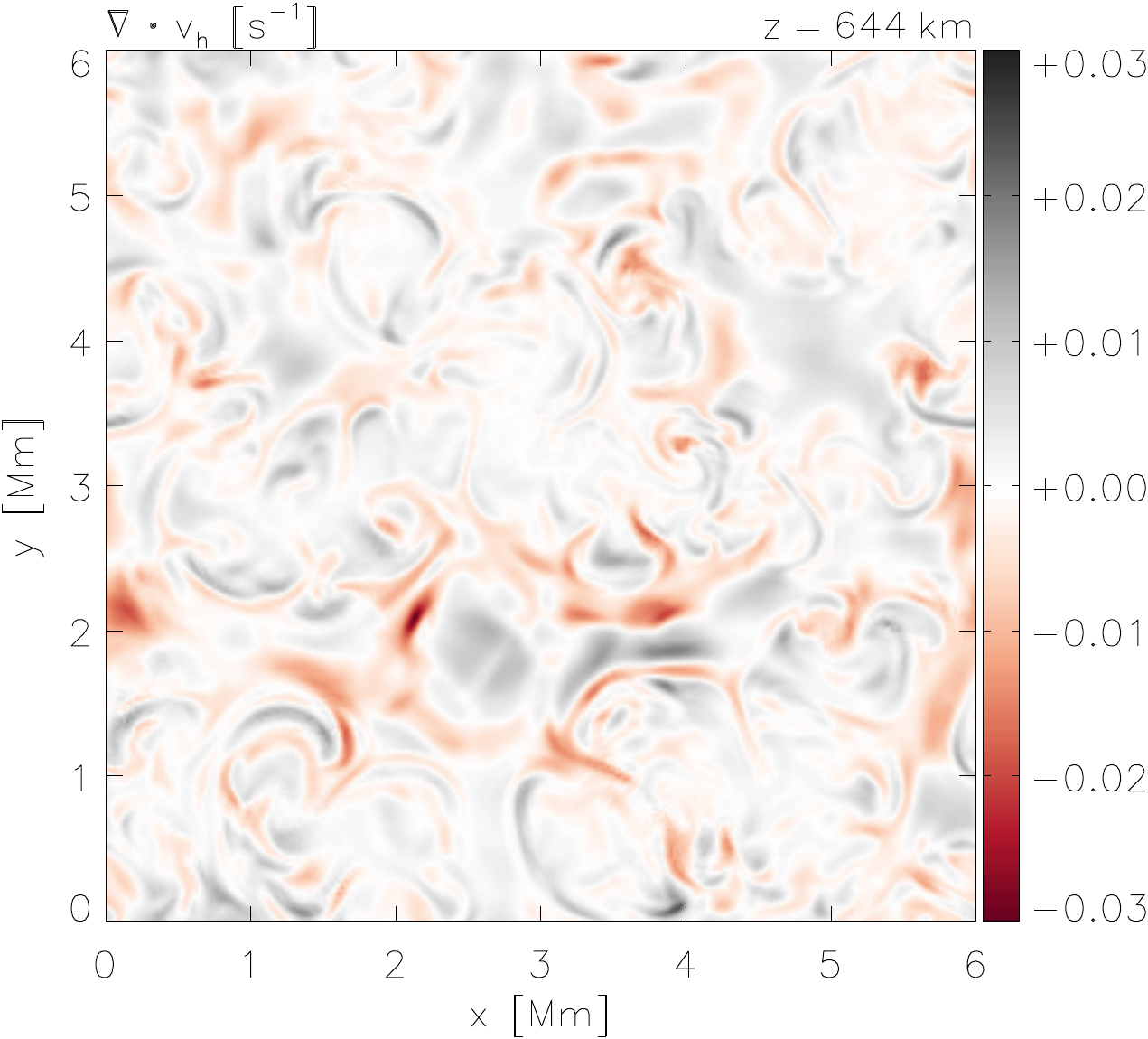}} 
          \\[\medskipamount]
{\includegraphics[scale=\myscale]{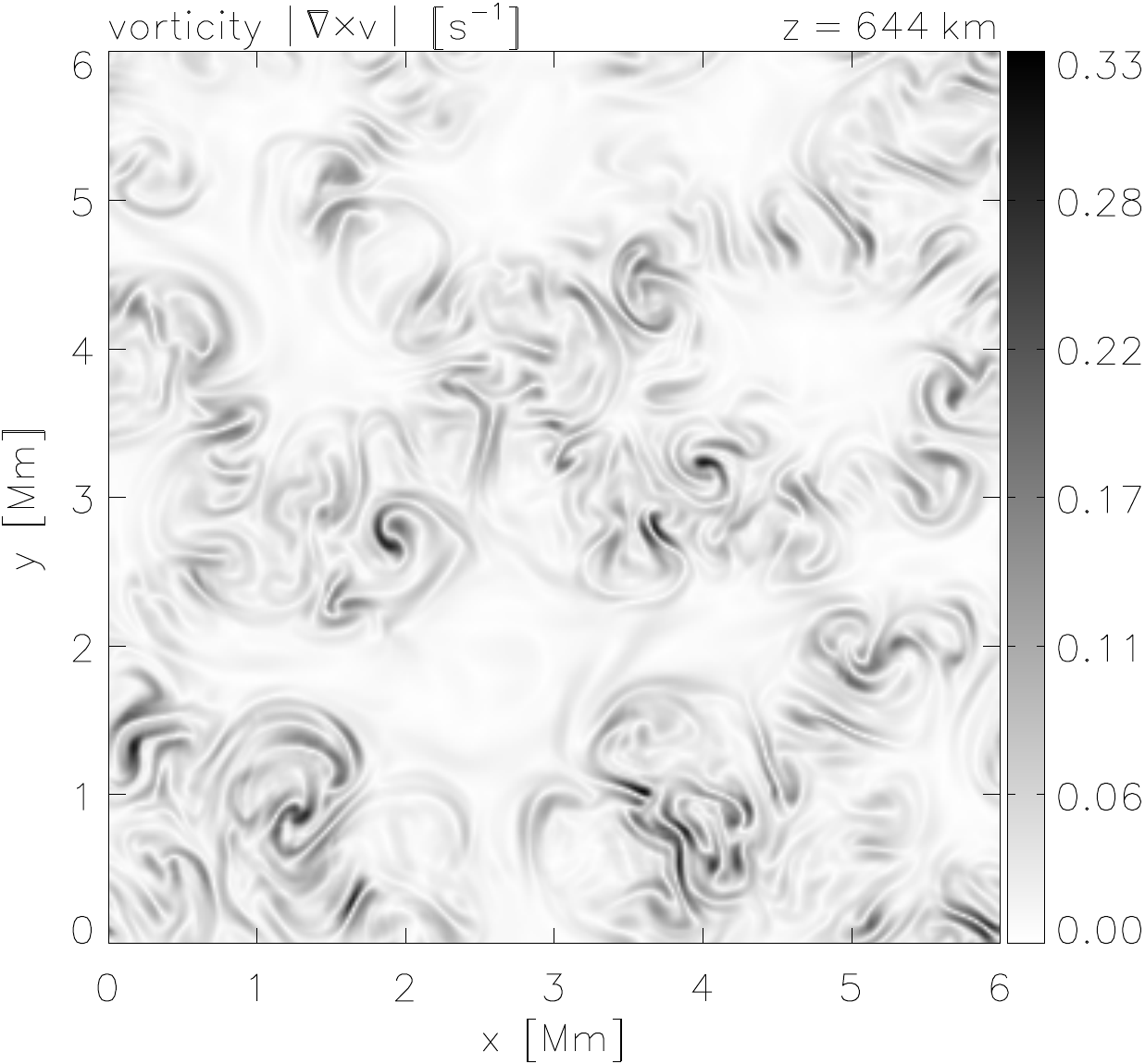}}\hglue 4mm
{\includegraphics[scale=\myscale]{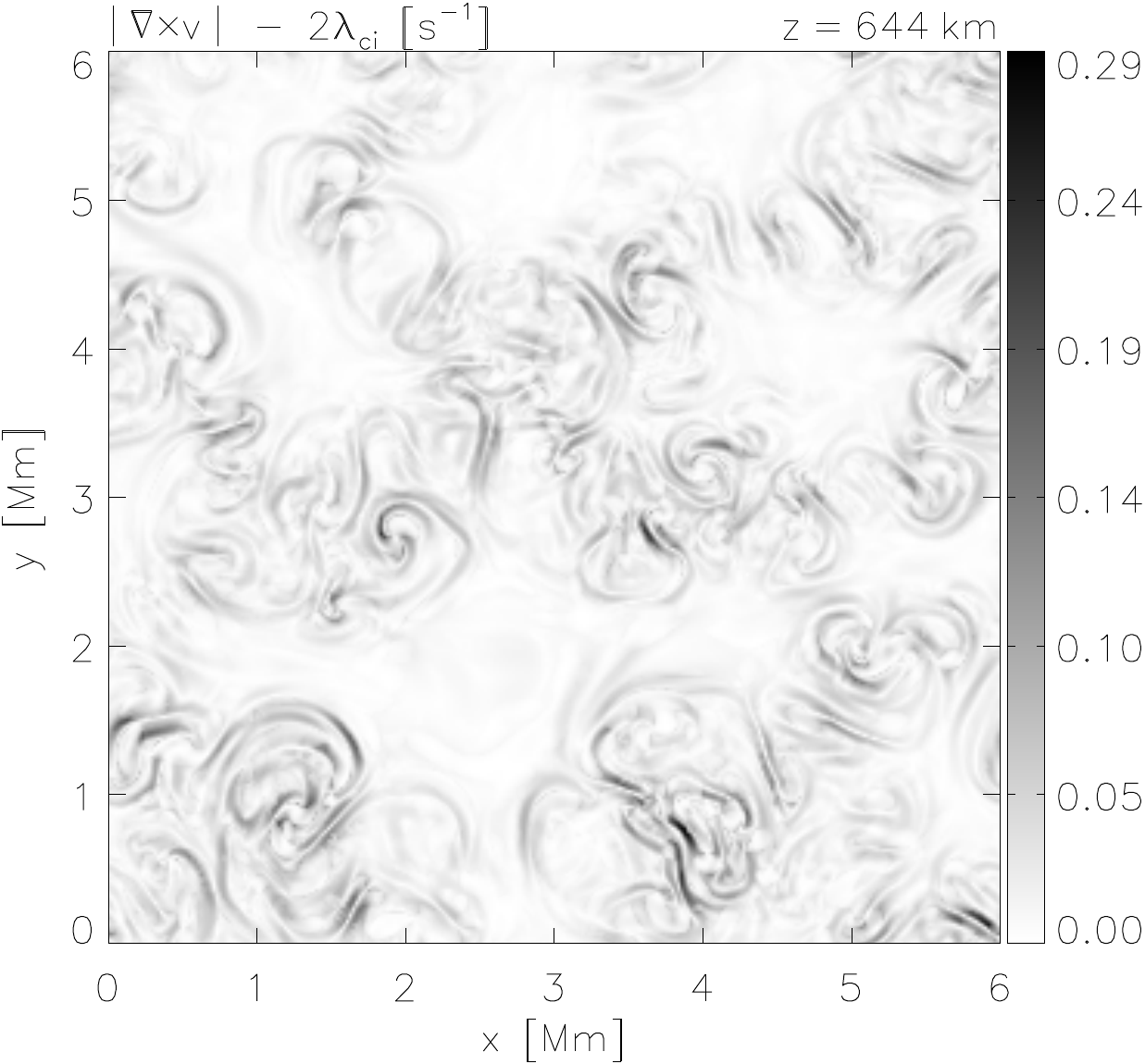}}
          \\[\medskipamount]
{\includegraphics[scale=\myscale]{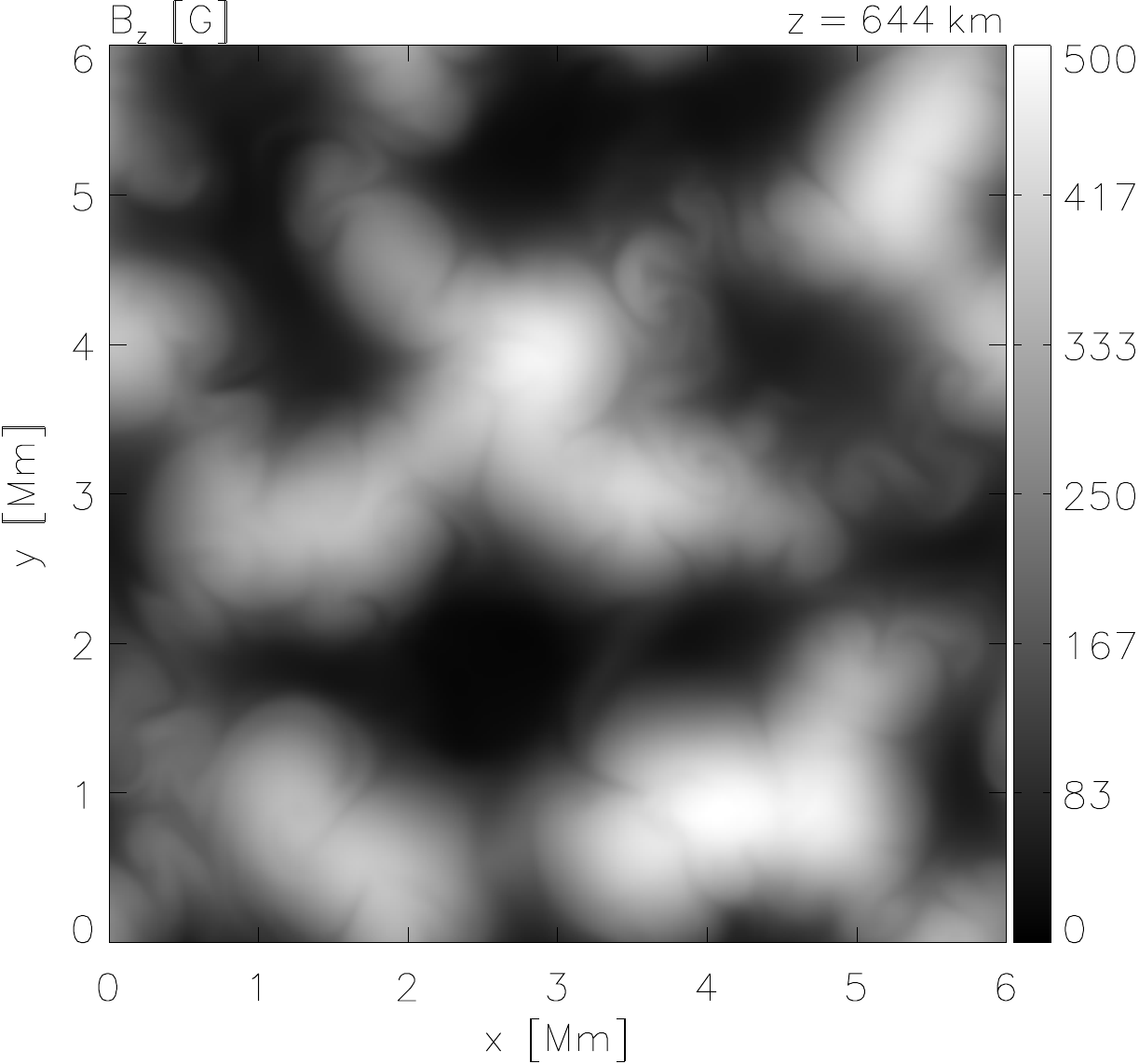}}\hglue 4mm
{\includegraphics[scale=\myscale]{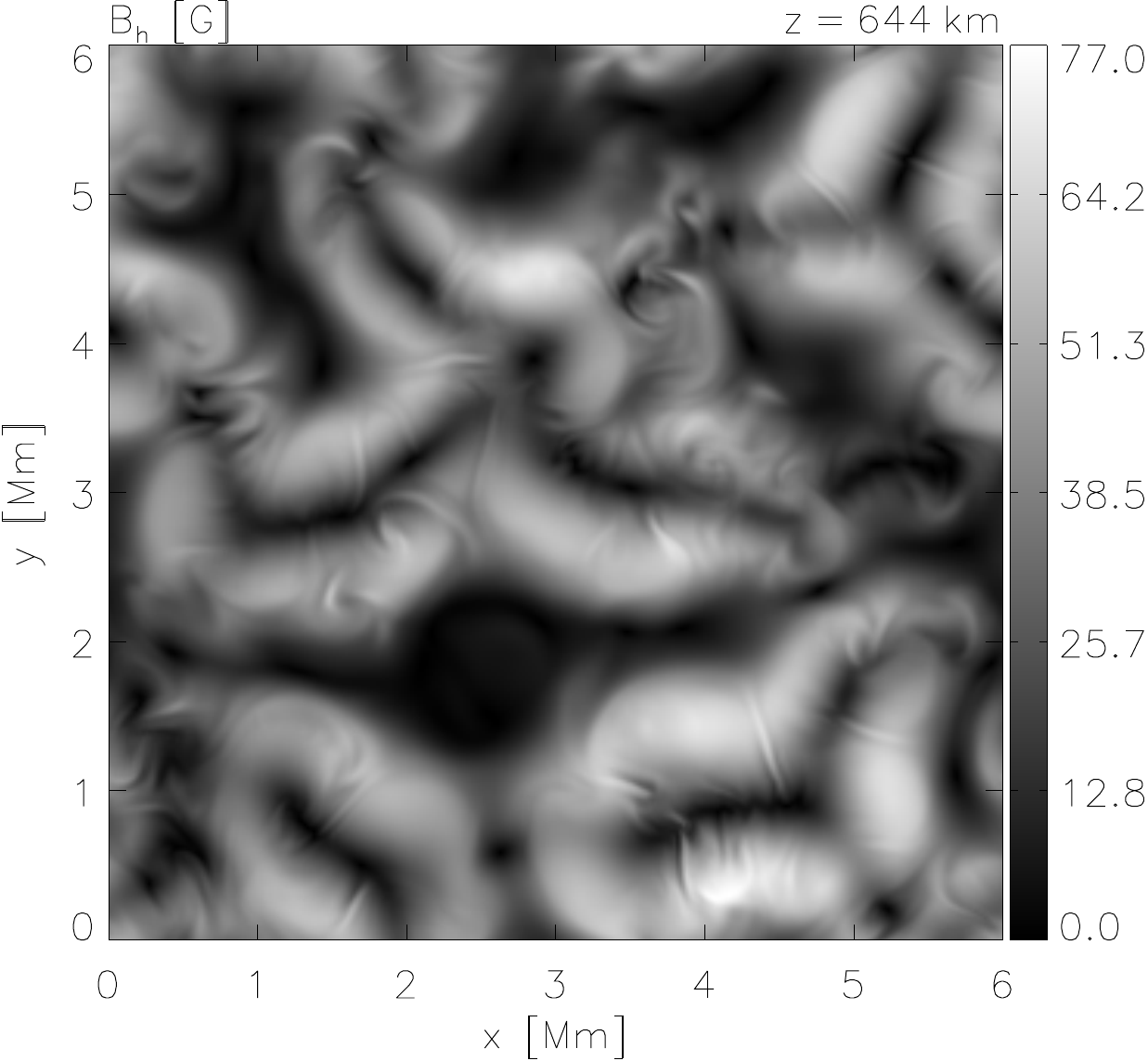}}
\caption{Horizontal maps of various quantities at $z=644\unit{km}$ for
  the magnetic case. {\em From left to right, top to bottom}:
  temperature, vertical velocity, swirling strength, divergence of the
  horizontal velocity, vorticity, shear part of vorticity, strength of
  the vertical magnetic field, and strength of the horizontal magnetic
  field.}
\label{fig:mapsB}
\end{figure*}

\subsection{Dynamics of the upper photosphere}
\label{subsec:dynamics}

High in the photosphere and above, the thermal and flow structure is
very different in the two cases.  Figs.~\ref{fig:mapsH}
and~\ref{fig:mapsB} show maps of various quantities on horizontal
planes. In the non-magnetic case (Fig.~\ref{fig:mapsH}), there is a
pattern of shock fronts with regions of strongly localized heating,
leading to a thermal bifurcation into hot and cool regions
\citep[cf.][]{Wedemeyer:etal:2004}. The generally inclined and
propagating shock fronts form a rapidly evolving filamentary pattern on
horizontal cuts of temperature and divergence of the horizontal
velocity. Some of the shocks are associated with vortices, which are
preferentially horizontally orientated (cf.~Fig.~\ref{fig:depthcov},
upper panel).  An example of a shock structure is given in
Fig.~\ref{fig:shock}, showing that the post-shock peak temperatures can
reach values around $7000\,$K, which exceed the temperature of the cool
background by about 3000~K.

In the magnetic case (Fig.~\ref{fig:mapsB}), the shock fronts are
virtually absent, the divergence of the horizontal velocities typically
being an order of magnitude smaller than in the non-magnetic simulation.
Apparently, the strong magnetic field suppresses the development of the
shock structures. Instead, a small-scale pattern of hot filaments
appears in the magnetic regions, which is tightly connected to
vertically orientated vortices and localized up- and downflows. The
high-temperature features are often associated with the shear part of
vorticity.

\begin{figure}[!ht]
\centering
\includegraphics[width=.9\linewidth]{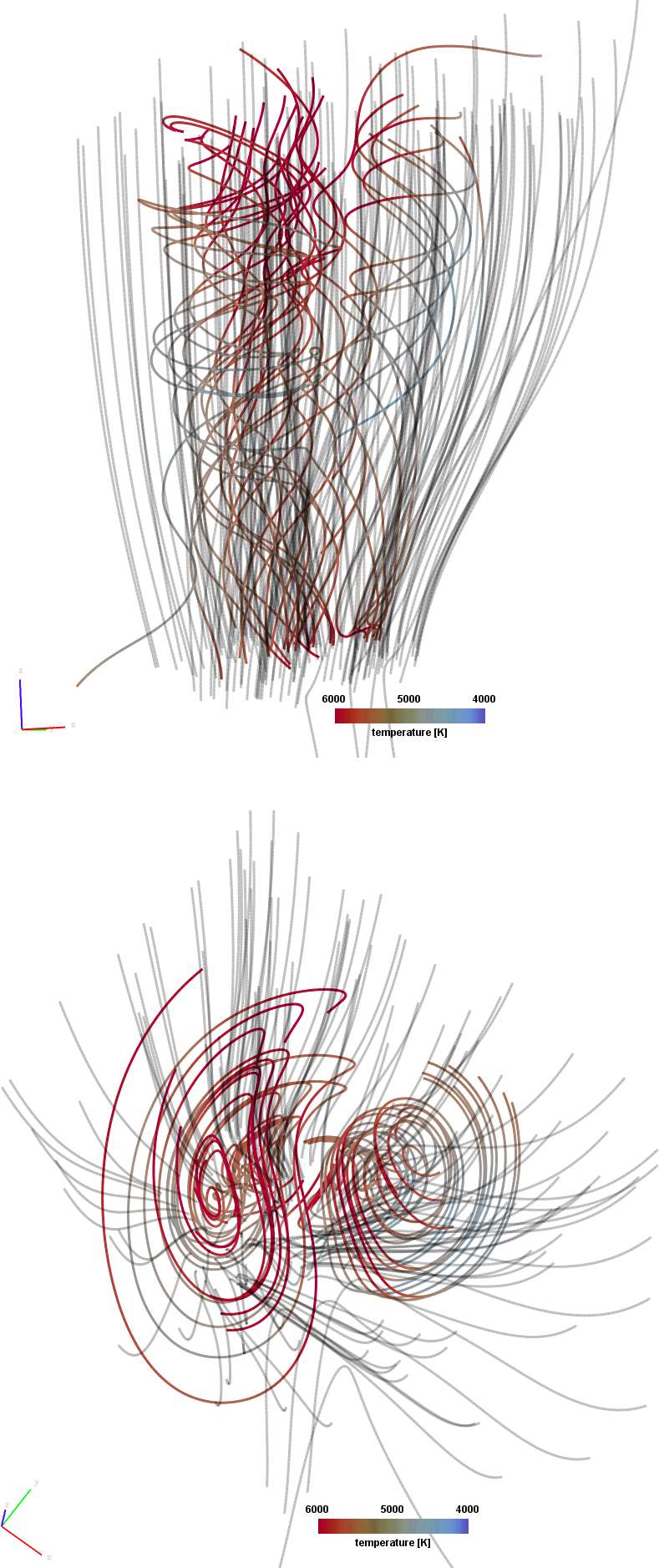}
\caption{Streamlines (colored) and magnetic field lines (gray) of two
 counter-rotating vortices above the optical surface. The vortex pair is
 located at $(x,y)\simeq(2.0\unit{Mm},2.9\unit{Mm})$ in
 Fig.~\ref{fig:mapsB}.  Color represents the temperature.}
\label{fig:dopstr}
\end{figure}

As an example of the flow structures in the magnetic simulation,
Fig.~\ref{fig:dopstr} depicts a pair of photospheric vortices in the
magnetic case.  The velocity streamlines are helically wound with a
pitch lower than the height of the photosphere. The magnetic field is
expanding with height and not significantly twisted. This is a
consequence of the high \Alfven speed in the photosphere, which reaches
about $70\unit{km\,s^{-1}}$ at the upper boundary of the computational
box: any twist of the vertical magnetic flux concentrations rapidly
escapes upward in the form of torsional \Alfven waves. Such processes
may drive swirling flows in the chromosphere
\citep{Wedemeyer:Rouppe:2009} and in the corona \citep{Zhang:Liu:2011}.

\subsection{Heating processes}

Figure~\ref{fig:lpdiss} shows height profiles of the temporarily and
horizontally averaged temperature and dissipation rates (viscous and
Ohmic). While the two simulations have very similar temperatures
in the deep and near-surface layers, the average atmosphere of the
magnetic run shows a much flatter temperature profile above $z\simeq
200\,$km.  At $z=600\,$km, the temperature difference
between the two runs reaches about 600~K. Above this height, the
non-magnetic atmosphere shows a steep temperature increase,
most likely due to shock heating. The flat profile in the magnetic case
indicates a different heating process that extends over a bigger
height range and affects also the middle photosphere.

In order to evaluate the importance of viscous and Ohmic dissipation as
heating processes, we determined the corresponding specific dissipation
rates, i.e., the change of internal energy per unit time and unit mass
due to viscous and magnetic diffusion, respectively.  The \MURaM
simulations analyzed here were carried out with a physical magnetic
diffusion term in the induction equation, assuming a constant magnetic
diffusivity of $\eta=1.1\times10^{11}\,$cm$^2\,$s$^{-1}$, the smallest value
compatible with the given spatial grid resolution.  In contrast, the
viscosity term in the momentum equation effectively acts only near the
grid scale. It involves artificial diffusivities explicitely depending
on the local velocity structure, namely hyperdiffusivities and
shock-resolving diffusivities \citep{Voegler:etal:2005}. For
stability reasons, there is also magnetic hyperdiffusion near the top and
bottom of the computational box, but this does not significantly
contribute to the dissipation rate below $z\simeq700\,$km.

The bottom panel of Fig.~\ref{fig:lpdiss} shows the height profiles of
the horizontally averaged specific dissipation rates for both runs.  The
viscous dissipation rate in the non-magnetic case rises steeply above
$z\simeq 600\,$km, probably related to shock formation in this height
range. In the magnetic simulation, the dissipation is dominated by Ohmic
dissipation below $z\simeq 500\,$km, which exceeds the viscous
dissipation by up to a factor of 5 in the middle photosphere. This
extended range of heating explains the relatively flat average
temperature profile in the magnetic case.  The continuous increase of
viscous dissipation above $z\simeq 200\,$km is partly a result of the
assumed height profile for the hyperdiffusivity, so that it is possibly
overestimated.

Figure~\ref{fig:dissipation} provides information about the spatial
distribution of the dissipation rates in the higher layers. In the
non-magnetic run, the main sources of viscous dissipation are 
shock fronts. In the magnetic case, the viscous dissipation%
\footnote{Since the hyperdiffusivities generally are anisotropic, the
  associated dissipation rates are not necessarily always positive.  A
  few grid cells indeed show slightly negative values of the viscous
  dissipation rate (see the small isolated white patches in the middle
  panel of Fig.~\ref{fig:dissipation}), but these are completely
  irrelevant for the distribution and the average of the dissipation
  rates.}
is associated with the vortices located in the magnetic flux
concentrations while the Ohmic dissipation rates are concentrated
at the edges of the magnetic flux concentrations.

More detailed and quantitative information on the relation between
temperature and flow structure in the upper layers is obtained from the
bivariate (2D) histograms shown in Fig.~\ref{fig:2Dhis}. The clearest
difference between the non-magnetic case (left column) and the
magnetic case (right column) is in the role of the velocity
divergence: while there is a strong correlation of negative divergence
(associated with shocks) with temperature in the non-magnetic case, such
a relation is almost absent in the magnetic run. In both cases, the
temperature correlates with swirling strength, but cells with moderate
to high swirling strength are much more abundant in the magnetic
case. In the non-magnetic simulation, the correlation with swirling
strength is due to the association of swirls with shock fronts.  Note
that the non-magnetic simulation has a much larger proportion of very
cool regions with temperatures between 2000~K and 3000~K, which are
almost absent in the magnetic case. Both simulations show cells with
high temperature for all values of the swirling strength, but
temperatures exceeding $7000\,$K apparently can only be provided by the
shock fronts in the non-magnetic case.

\renewcommand{\myscale}{.45}
\begin{figure}[ht!]
\centering
{\includegraphics[scale=\myscale]{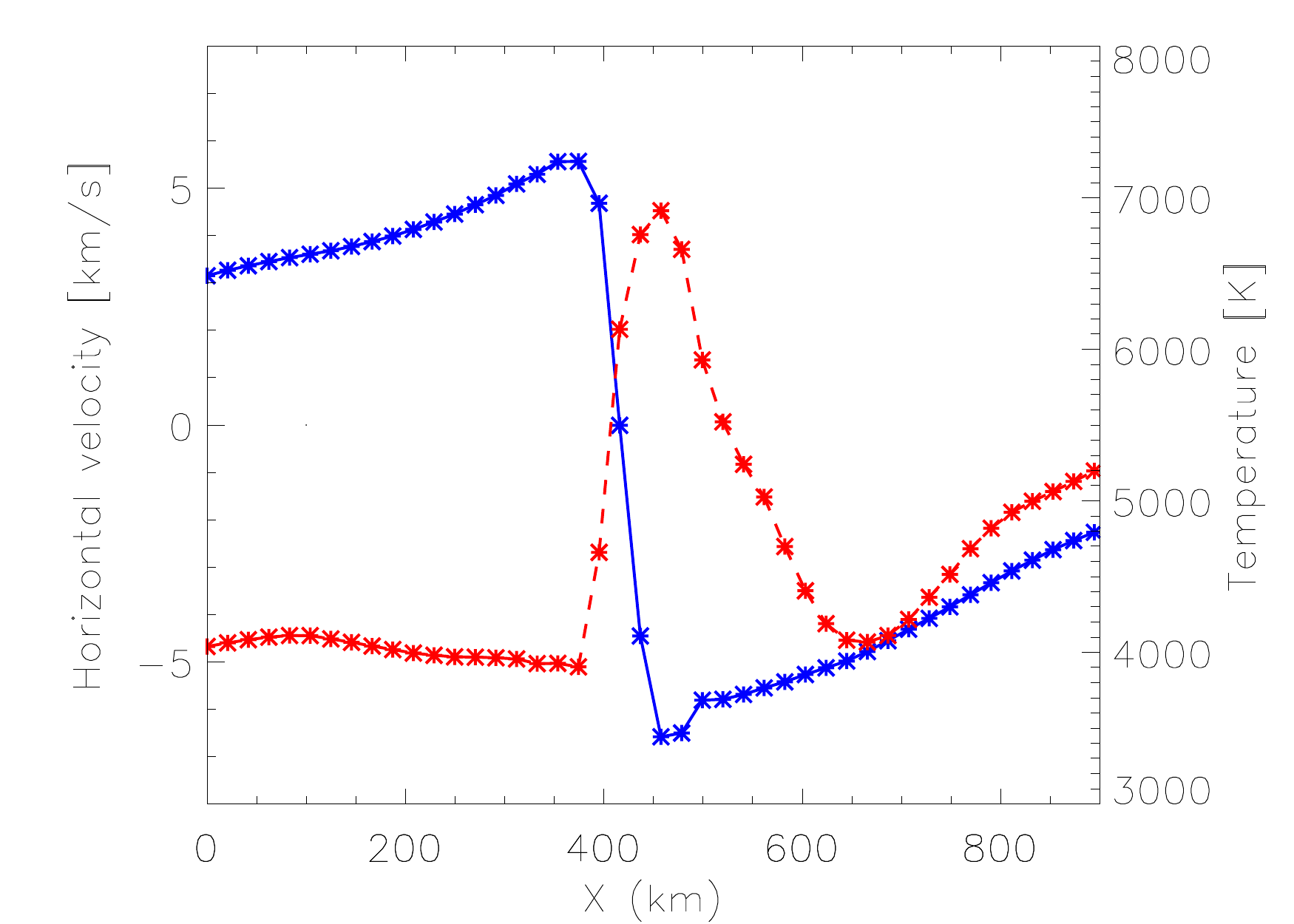}} 
{\includegraphics[scale=\myscale]{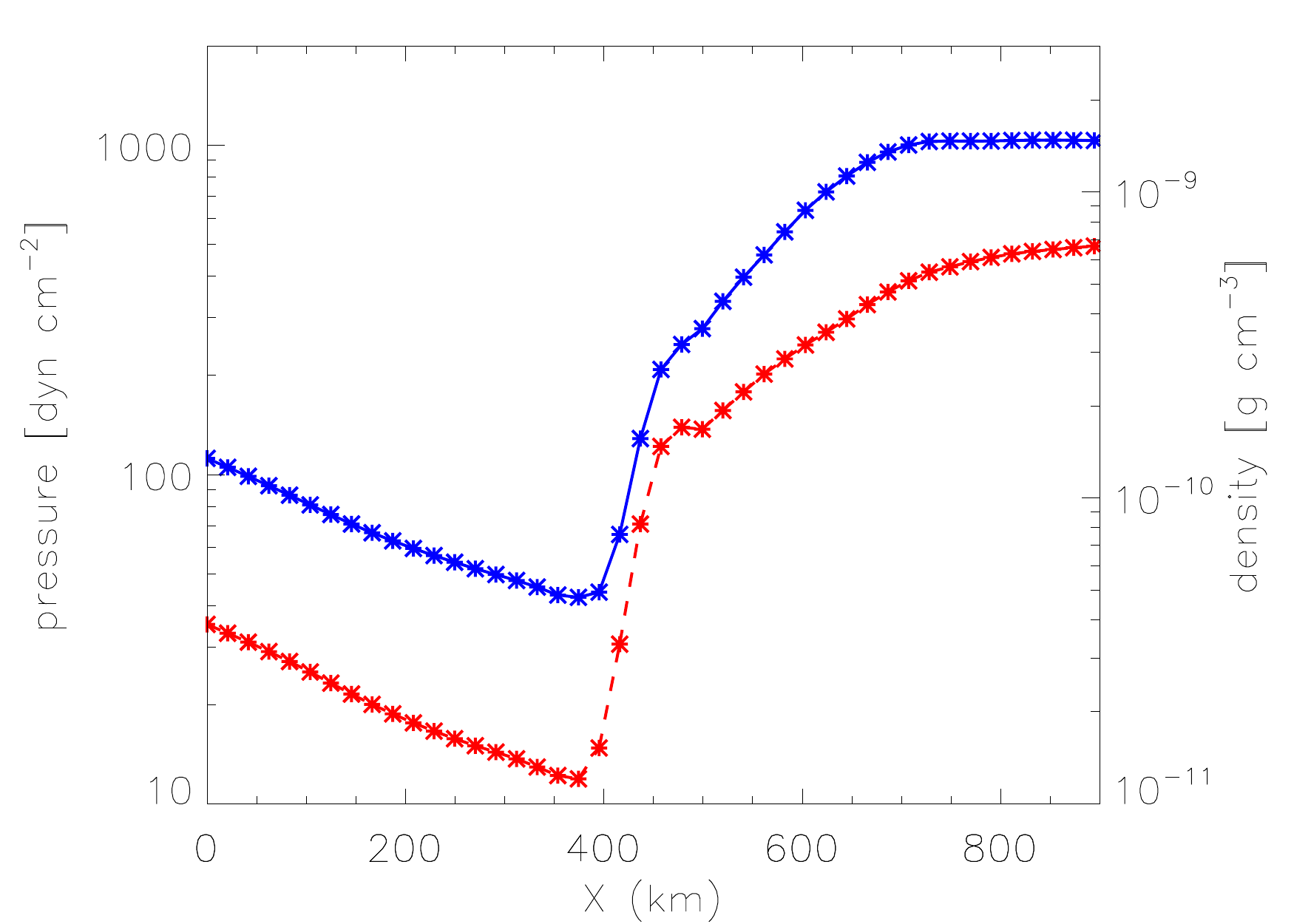}} 
\caption{Horizontal cuts in $x$-direction through the shock located at
  $(x,y)=(4.5\unit{Mm},5.7\unit{Mm})$ in Fig.~\ref{fig:mapsH}. Upper
  panel: horizontal velocity (blue, solid line) and temperature (red,
  dashed line); lower panel: density (blue, solid line) and gas pressure
  (red, dashed line).}
\label{fig:shock}
\end{figure}

\section{Summary and discussion}
\label{sec:discussion}

The simulation results reveal a striking difference between the magnetic
and the non-magnetic case in terms of the velocity structure in the
upper photosphere and around the temperature-minimum region. While these
layers are dominated by shocks in the non-magnetic simulation, the
magnetic case is characterized by extended vertical vortices associated
with the magnetic flux concentrations.  Clearly, there must be a
transition from the shock-dominated non-magnetic regime to the
vortex-dominated magnetic regime.  High-resolution simulations of
small-scale dynamo action \citep{Graham:etal:2010} indicate that the
shock pattern is almost unaffected by the low-lying magnetic loops and
occasionally forming vertical flux concentrations generated by the
small-scale dynamo.  A run simulating the decay of a mixed-polarity
field \citep[cf.][]{Cameron:etal:2011} shows that the suppression of the
shock pattern sets in for a mean strength of the vertical field
component of about 20~G in the lower photosphere. This is consistent
with the simulations of \citet{Schaffenberger:etal:2006}, who find a
clear pattern of shock waves for a mean vertical field of 10~G.
Consequently, `quiet' regions with a mean unsigned vertical field around
10~G \citep[e.g.,][]{Lites:etal:2008} are probably dominated by shock
waves, while the upper photosphere of more strongly magnetized unipolar
regions (in network and plage) is dynamically dominated by vortices and
shear flows associated with the magnetic features.

In both simulations, substantial local heating takes place: shock
heating in the top layers of the non-magnetic case and a combination of
Ohmic heating (in the lower to middle photosphere) and viscous heating
(in the upper photosphere) in the magnetic case. The different height
variation of the heating rates in the two cases leads to a sharp
temperature rise in the upper layers of the non-magnetic case, in
contrast to a more extended temperature enhancement resulting in a
flatter temperature gradient in the magnetic run.

\renewcommand{\myscale}{.5}
\begin{figure}[ht!]
\centering
\includegraphics[scale=\myscale]{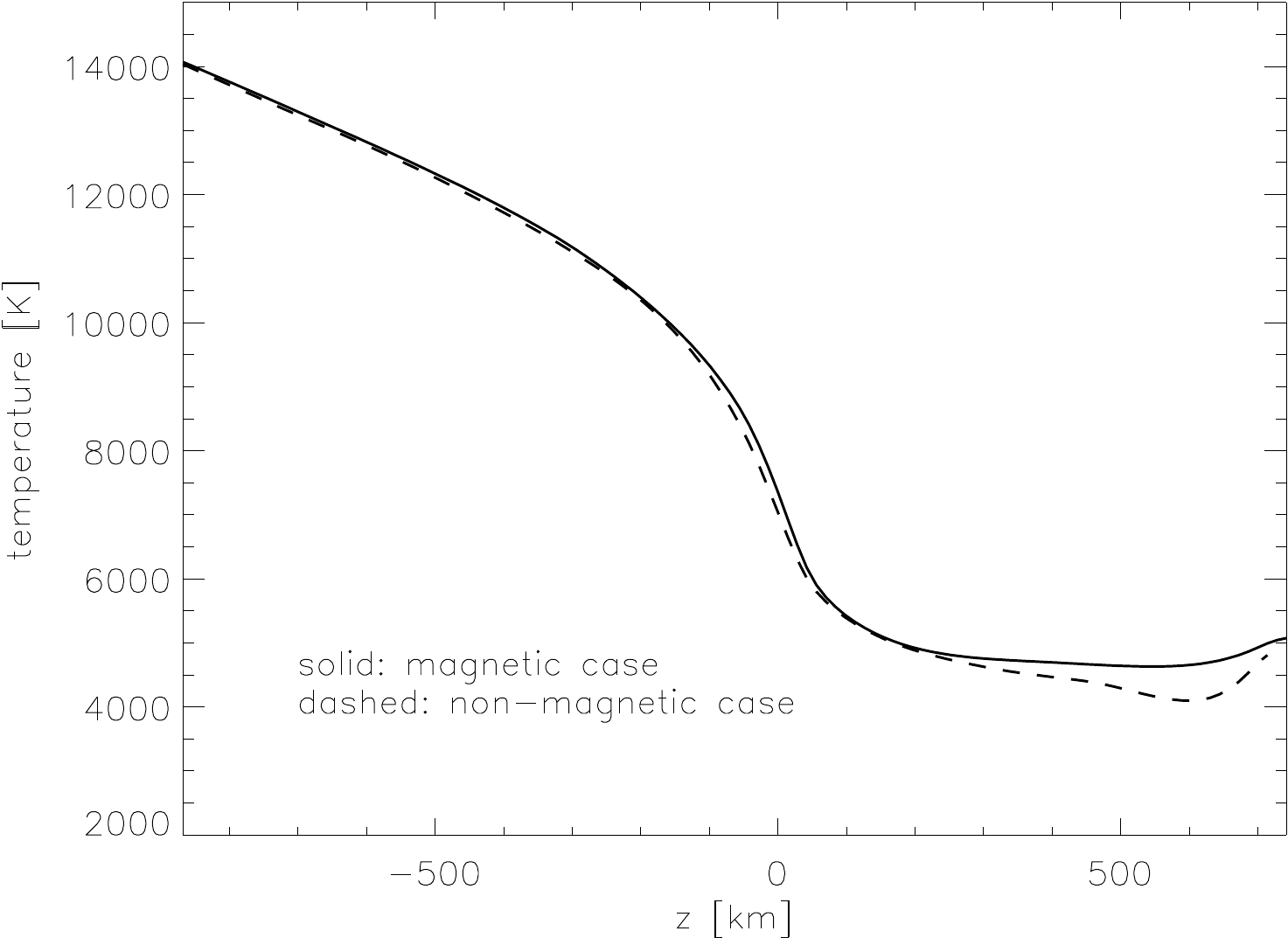}
\vglue 3mm
\includegraphics[scale=\myscale]{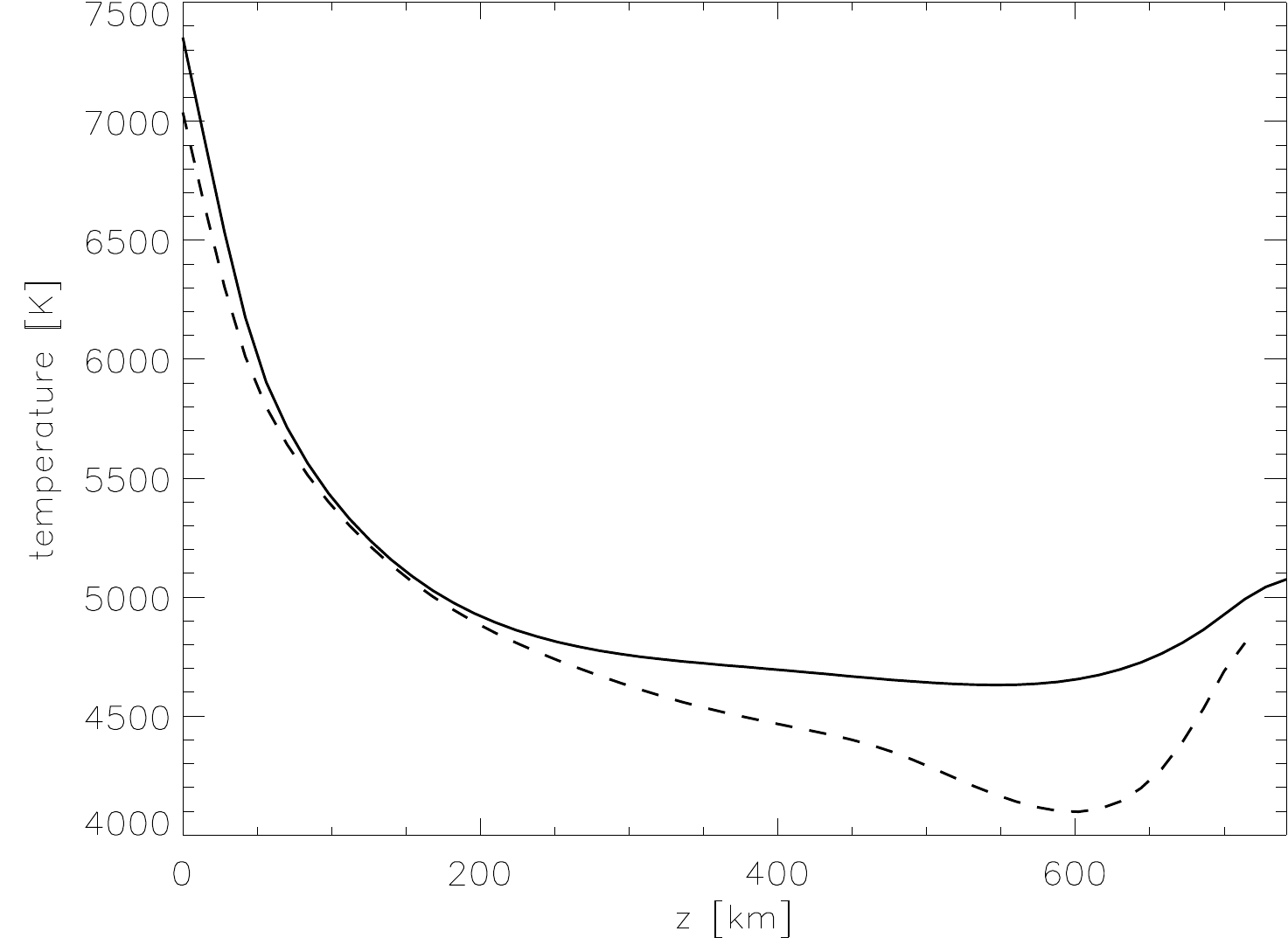} 
\vglue 3mm
\includegraphics[scale=\myscale]{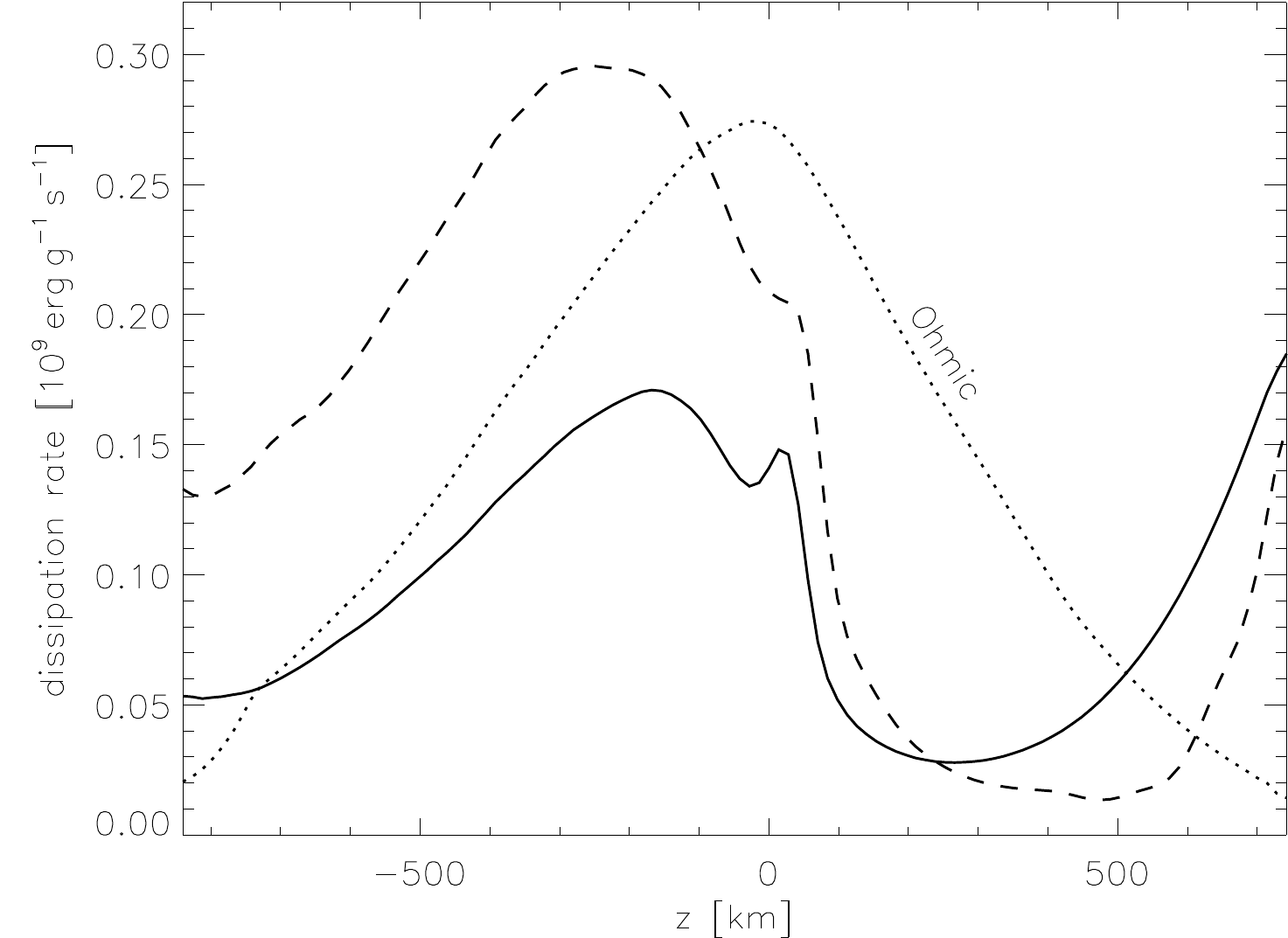}
\caption{Height profiles of the horizontally averaged temperature
  over the full height range (top panel) and above the optical
  surface (middle panel). The bottom panel shows the horizontally
  averaged  viscous dissipation rates (solid: magnetic case; dashed:
  non-magnetic case) and the Ohmic dissipation rate in the magnetic case
  (dotted). }
\label{fig:lpdiss}
\end{figure}

\renewcommand{\myscale}{.5}
\begin{figure}[ht!]
\centering
\includegraphics[scale=\myscale]{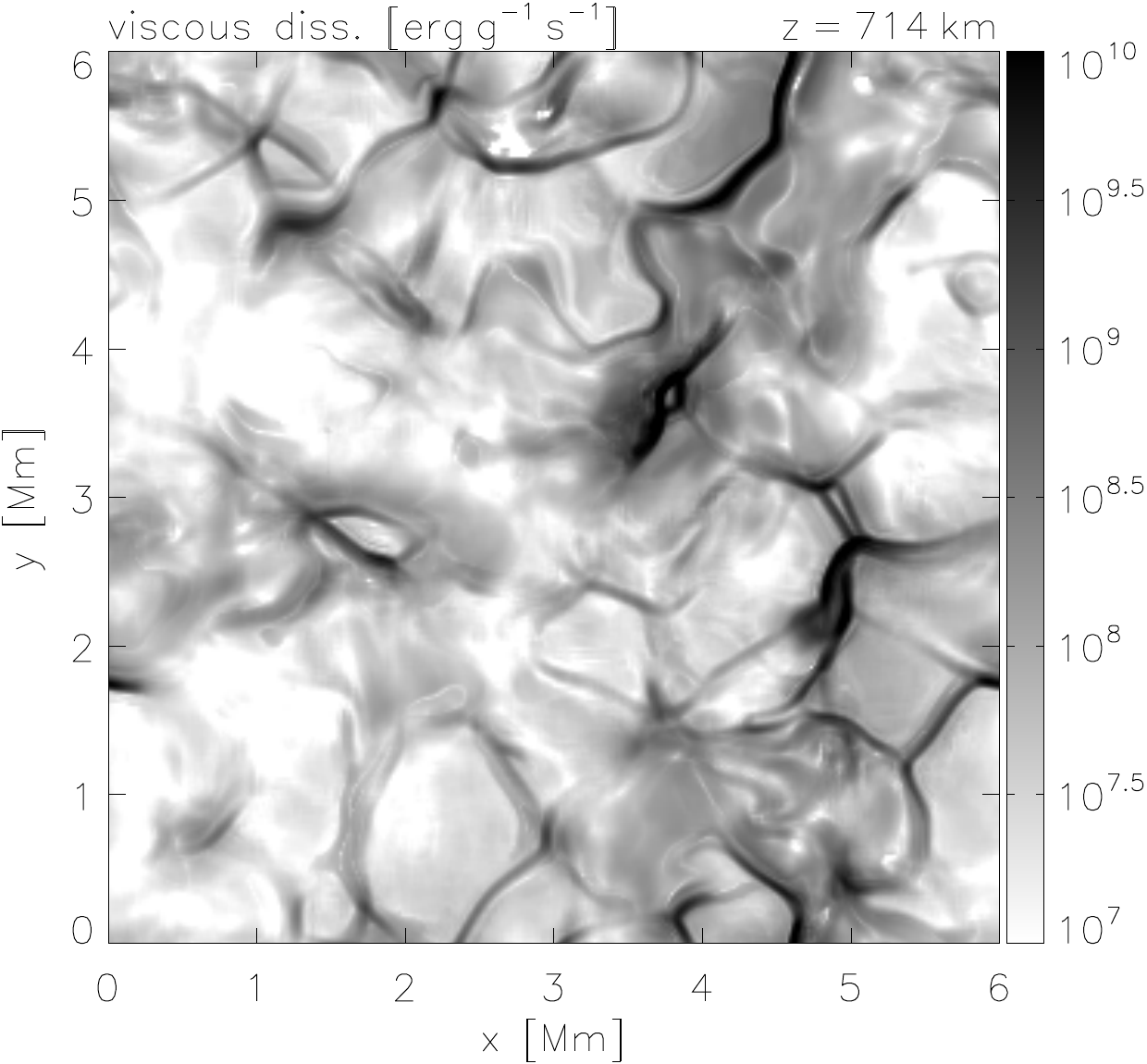}\\
\includegraphics[scale=\myscale]{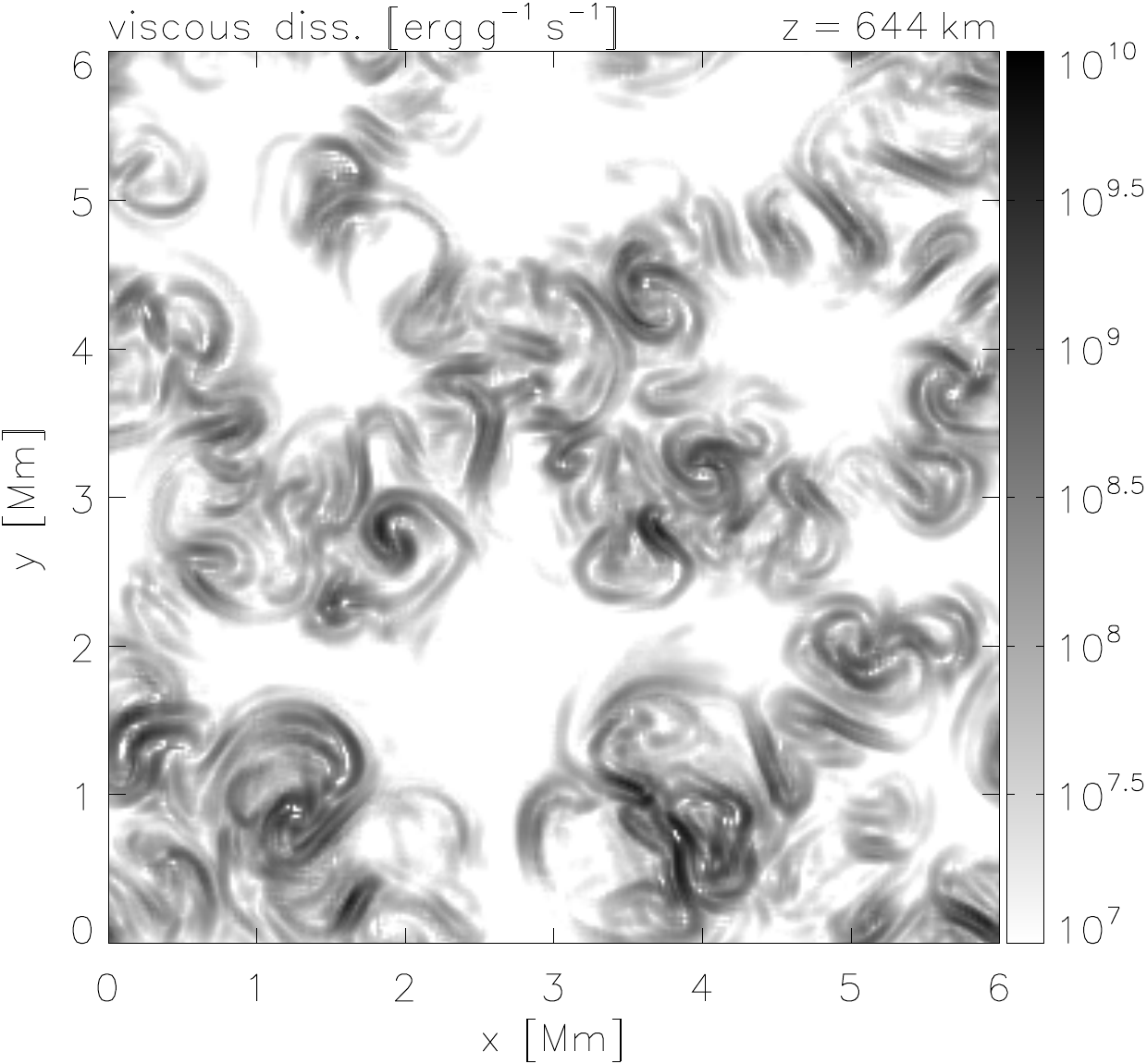}\\
\includegraphics[scale=\myscale]{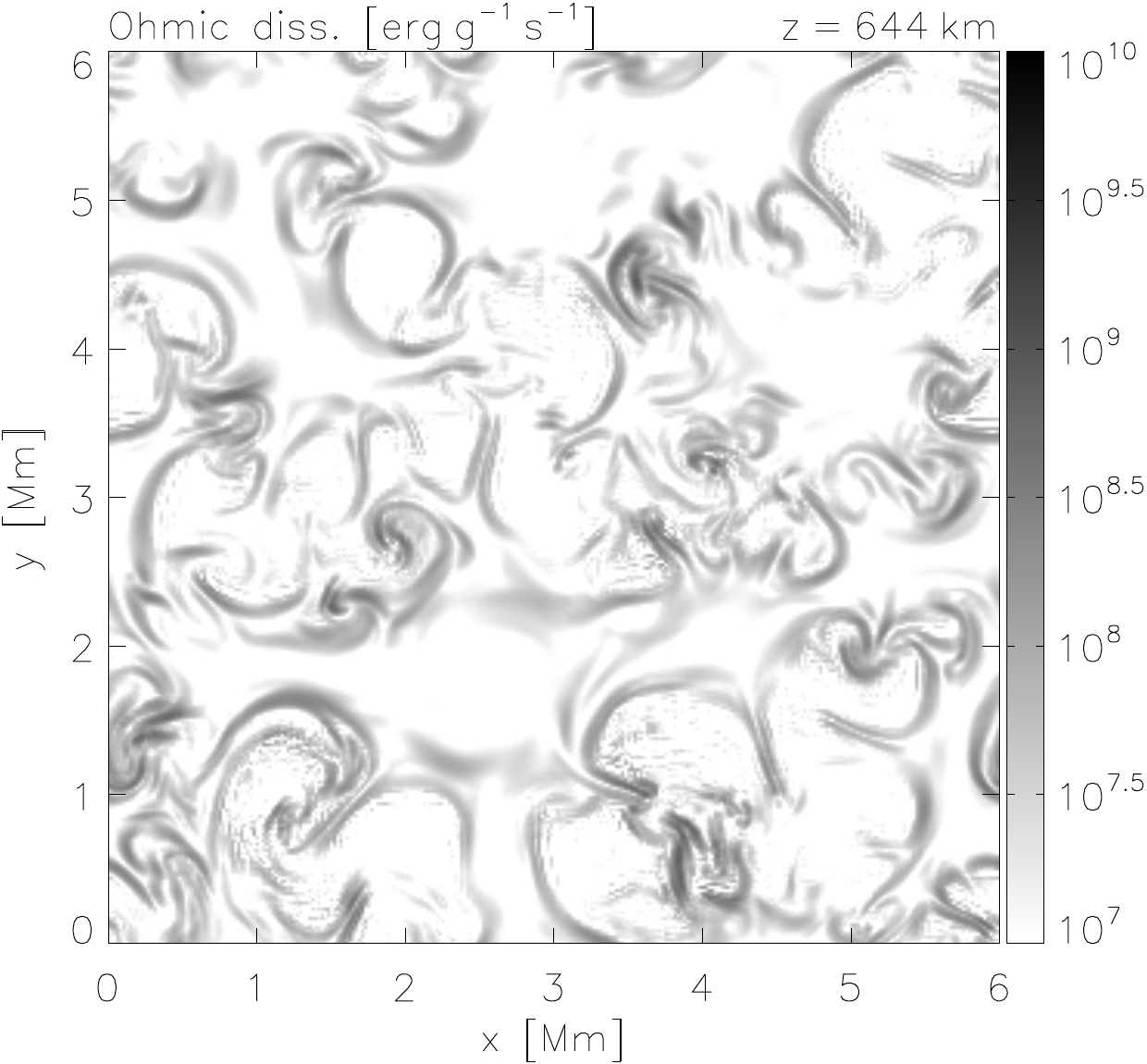} 
\caption{Dissipation rates on horizontal cuts through the upper
photosphere. Top: viscous dissipation in the non-magnetic case; middle:
viscous dissipation in the magnetic case; bottom: Ohmic dissipation. }
\label{fig:dissipation}
\end{figure}

However, while yielding important qualitative insight in the different
mechanisms for mechanical heating, we do not consider the dissipation
rates and thus the temperature structure in the higher layers provided
by the simulations to be quantitatively reliable. In the simulation runs
considered here, the hyperdiffusive viscous dissipation rate was
arbitrarily enhanced in the upper photosphere for reasons of numerical
stability, which affects the temperatures.  It is
also not clear how strongly the dissipation rates depend on the spatial
grid resolution and on the assumption of a vertical field at the upper
boundary (as opposed to matching to a potential field, for
instance). Apart from these problems, we have to keep in mind that the
conditions in the layers around the temperature minimum probably deviate
from the gray LTE radiative transfer used in the
simulations. Furthermore, a direct comparison between horizontal
temperature averages (or even averages over surfaces of constant optical
depth) from simulations with semi-empirical models would be problematic
owing to the strong horizontal inhomogeneities and the nonlinear
averaging underlying observed spectral mean line profiles
\citep{Uitenbroek:Criscuoli:2011}.

We intend to continue this study by carrying out new simulations with an
improved treatment of the subgrid scales \citep[see][]{Rempel:etal:2009}
and by performing a detailed investigation of the dependence of the
dissipation rates on grid resolution and on the upper boundary condition
for the magnetic field.

\begin{acknowledgements}
This work has been supported by the Max Planck Society in the framework
of the Interinstitutional Research Initiative \textit{Turbulent
transport and ion heating, reconnection and electron acceleration in
solar and fusion plasmas} of the MPI for Solar System Research,
Katlenburg-Lindau, and the Institute for Plasma Physics, Garching
(project MIF-IF-A-AERO8047).
\end{acknowledgements}

\renewcommand{\myscale}{.45}
\begin{figure*}[ht!]
\centering
\includegraphics[scale=\myscale]{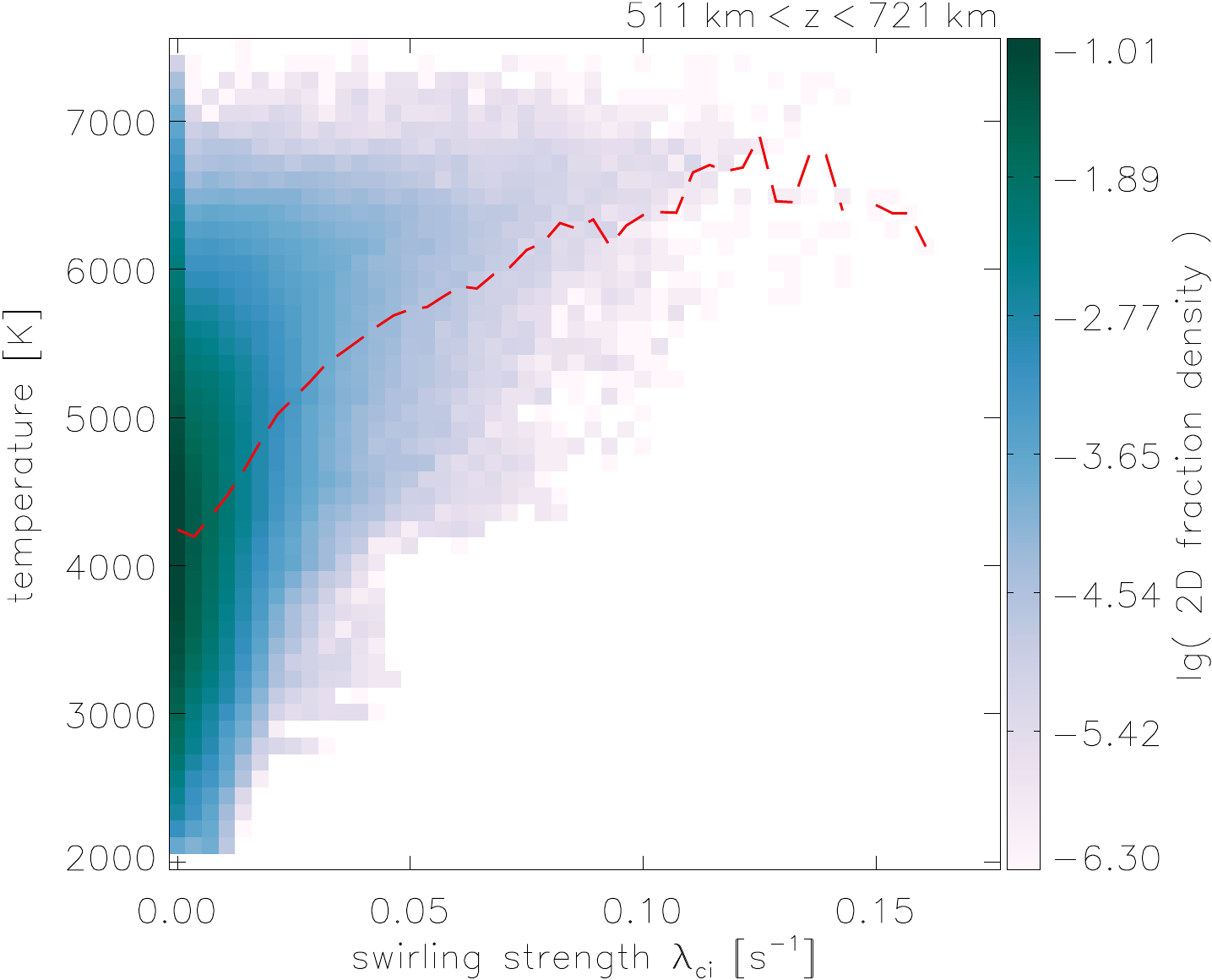}\hglue 4mm 
\includegraphics[scale=\myscale]{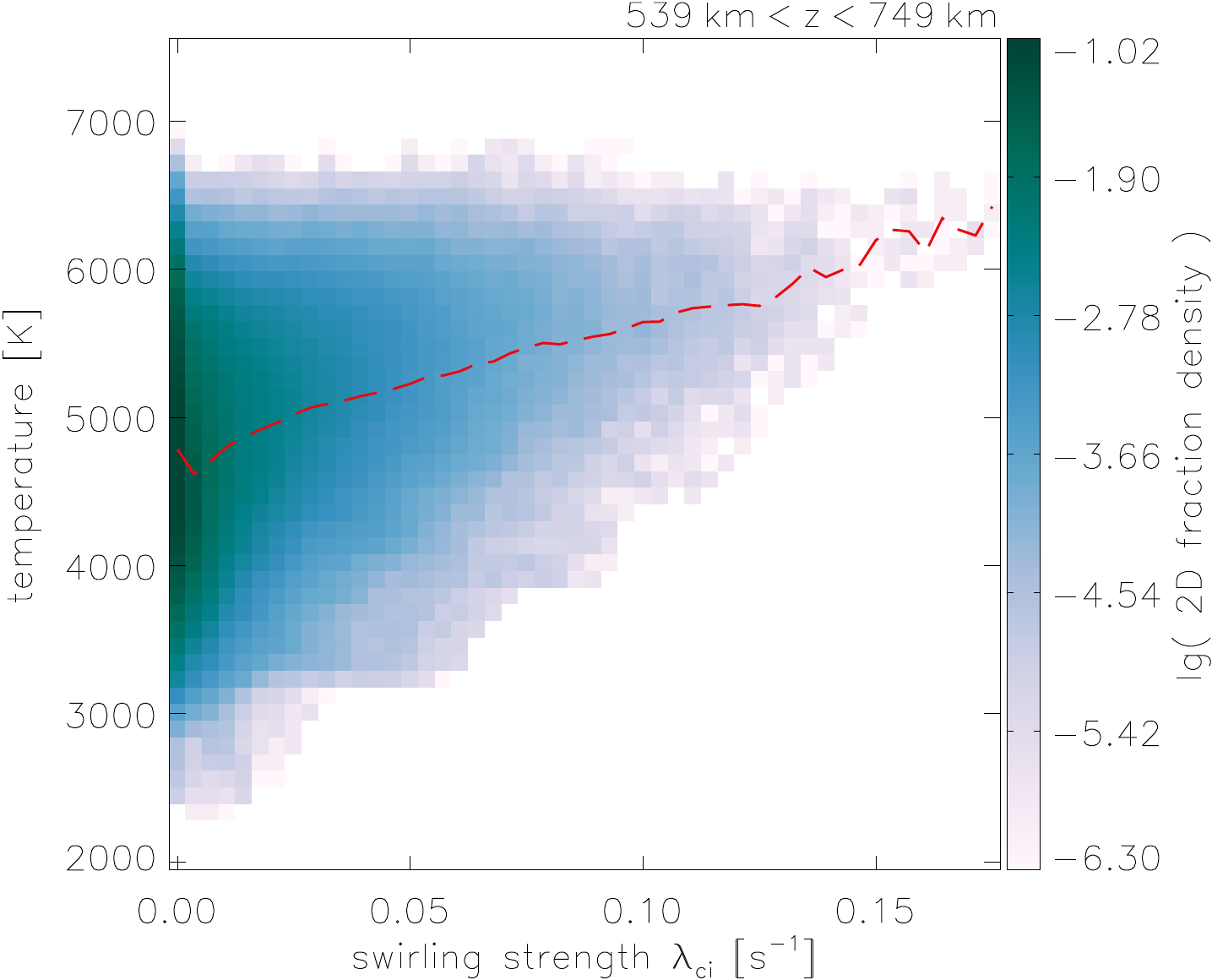} \\
[\medskipamount]
\includegraphics[scale=\myscale]{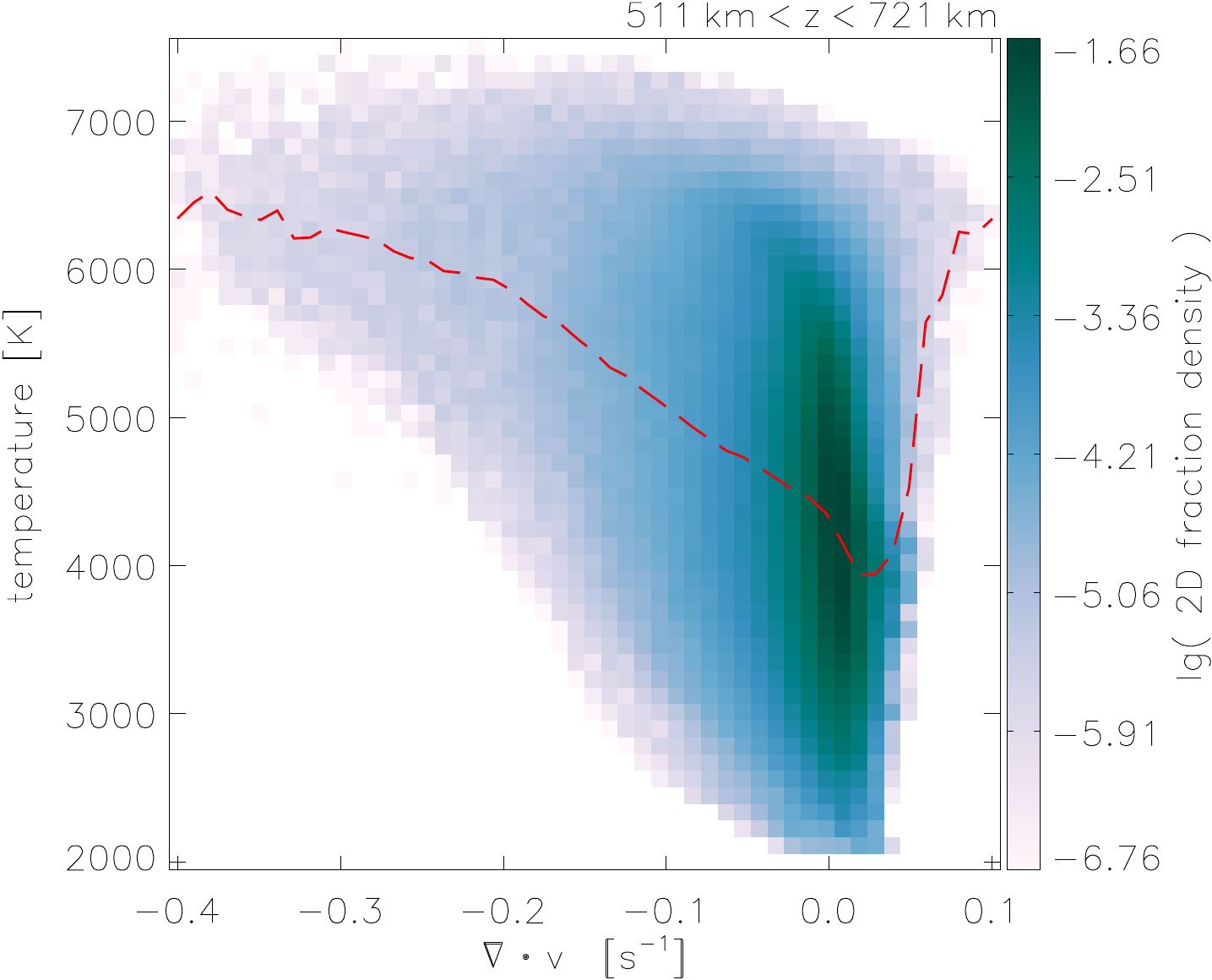}\hglue 4mm 
\includegraphics[scale=\myscale]{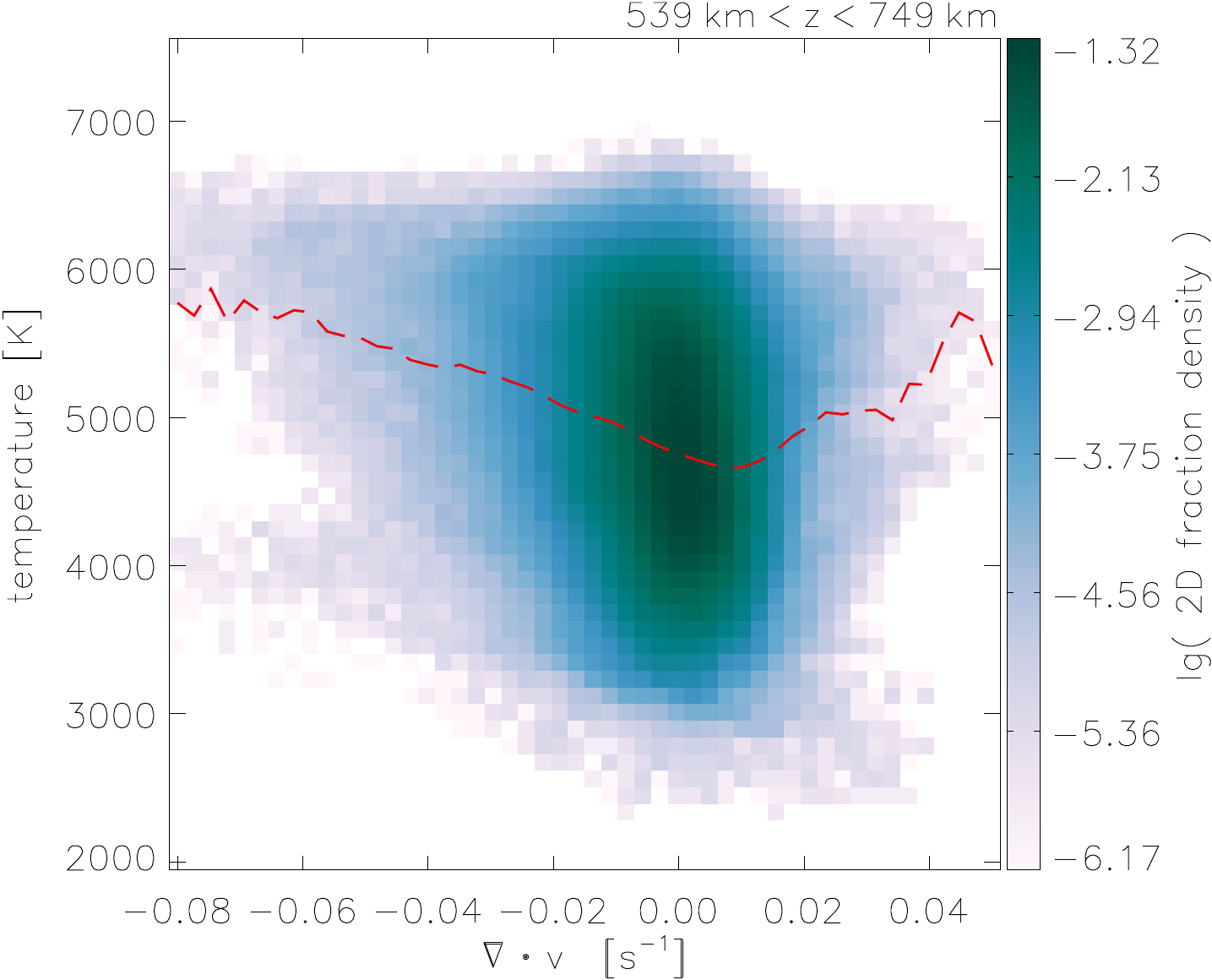}
\caption{2D histograms relating the temperature (vertical
axes) to swirling strength (upper panels) and velocity divergence (lower
panels) in the upper photosphere. Results for the non-magnetic case are
shown on the left panels and those for the magnetic case on the right
panels. The dashed red lines indicate the mean temperature at a given
value of the quantity on the horizontal axis. Each histogram is
normalized by its 2D integral.}
\label{fig:2Dhis}
\end{figure*}

\bibliography{joushort,aa18866}

\end{document}